\documentclass[letterpaper,twocolumn,10pt]{article}
\usepackage{usenix}
\AtBeginDocument{}
\usepackage{graphicx}
\usepackage{subcaption}
\usepackage{tikz}
\usetikzlibrary{arrows.meta, positioning, shapes.geometric, calc, fit, backgrounds}
\usepackage{booktabs}
\usepackage{multirow}
\usepackage{url}
\usepackage{amsmath, amssymb}
\newcommand{\rev}[1]{#1}                       

\begin{document}

\title{LoRA as Oracle}

\author{
	{\rm Marco Arazzi}\\
	University of Pavia\\
	marco.arazzi01@universitadipavia.it
	\and
	{\rm Antonino Nocera}\\
	University of Pavia\\
	antonino.nocera@unipv.it
}
\date{}
\maketitle

\begin{abstract}
Practitioners increasingly deploy neural networks they did not train, and must audit them after the fact for hidden backdoors, without the training pipeline, the poisoned data, or knowledge of any trigger. We introduce a low-rank auditing lens built on a single observation: what a model has internalized and how it behaves are distinct axes that can diverge. Fitting a small low-rank adapter toward a hypothesis and reading the geometry of the resulting update, its energy relative to, and its alignment with, the frozen weights, measures internalization directly, independently of the model's output behavior.
The lens earns its value where the two axes diverge: a backdoor is malicious internalization, a learned trigger-to-target shortcut, that behavioral auditing can miss. Reading it, we identify the backdoor's target class label-free and without any triggered data when the backdoor leaves a legible internalization signature, then erase the identified shortcut within the same low-rank subspace, uniquely coupling detection confidence to the size of the repair. Across four datasets and four architectures, our lens is the most consistent target identifier among defenses in its threat model; its rank-$r$ repair removes backdoors while preserving clean accuracy where full-model baselines collapse it, at orders-of-magnitude lower parameter and memory cost; and it is the only method that both erases the backdoor and leaves a benign model intact, because it can decline to act when uncertain.
\end{abstract}

\section{Introduction}

Most of the existing AI-based solutions make use of large models that are typically pre-trained on massive and heterogeneous corpora of data. The complexity of the adopted data makes it very difficult to certify its composition and security, raising important concerns about the quality of training and the reliability of the model obtained. In particular, the most relevant aspects to consider are related to whether specific data sources were used during training and whether malicious manipulations occurred in the learning process. These issues are even more critical in scenarios where the adopted models are open-source or shared among entities and deployed inside the boundaries of organizations with limited or no knowledge on the training process.

Consequently, there is a pressing need for auditing and security solutions that are both computationally efficient and capable of capturing global signals related to data usage and training-time threat injections and model manipulations. Regarding the first aspect, contributions verifying whether specific data points were included in the training process of a target model are commonly referred to as Membership Inference Attack (MIA) approaches \cite{shokri2017membership,yeom2018privacy}. However, most of these approaches typically rely on extensive querying, auxiliary shadow models, or strong assumptions about the original training set and access to the target model \cite{hu2022membership}; therefore, they are poorly suited to generic deployment scenarios. Concerning the second aspect, a large proportion of proposed approaches \cite{abad2025sok} target scenarios in which defenses can be deployed during the training phase \cite{wei2024mitigating,guo2025prototype}, while approaches focusing on post-training detection of anomalies typically require either access to at least a portion of the original training data, or specialized detection pipelines that analyze the internal activations of the model \cite{chen2018detecting}. However, obtaining access to even a small fraction of the training data is often infeasible, due to both practical constraints and, above all, privacy restrictions. Moreover, as target models increase in complexity, approaches relying on in-depth analysis of internal components become increasingly impractical.

Targeting the aforementioned gaps in the current reference literature, in this work, we consider a defensive auditing setting in which a data contributor or model user seeks to verify properties of a pre-trained model. In particular, the objective of the audit is twofold: (i) to determine whether a specific dataset was used during pretraining, and (ii) to detect the presence of a hidden threats (such as a backdoor) in the model.
We repurpose parameter-efficient fine-tuning via Low-Rank Adapters (LoRA), originally an efficiency tool, as a probing mechanism of the model's internal representations, avoiding exhaustive internal inspection or retraining of the target model.
In practice, we develop a lightweight framework, namely LoRAcle, that acts as a lens into a target model by fitting low-rank adapters on top of it and analyzing the geometry of the resulting updates. Our central observation is a single principle: what a model has internalized and how it behaves are distinct axes, and they can diverge. A low-rank adapter trained on content already internalized by the original model requires only a small, well-aligned update, whereas content that the original model can merely process but has not internalized demands a larger and less aligned adaptation. This geometric signal is not the same as the model's output behavior, and the gap between the two is exactly where auditing lives: a backdoor can be behaviorally loud yet geometrically faint, or behaviorally quiet yet geometrically obvious. Reading this internalization axis through simple ranking and energy-based statistics, without access to the original training data or shadow models, enables us to detect and remove backdoors which are missed by behavior-based auditing. We first validate this lens on membership inference, where the behavioral and internalization axes coincide, before applying it to backdoor detection, where they diverge.
In summary, the contributions of this work are as follows:
\begin{itemize}
    \item We introduce a low-rank auditing lens that audit the knowledge internalized by frozen models studying the geometry of rank-$r$ adapter updates, and validate this signal through membership inference.
    \item We apply this lens to label- and trigger-free backdoor target identification, matching existing detectors under equivalent threat models.
    \item We exploit low-rank accessibility to remove backdoors through rank-$r$ unlearning, coupling detection confidence directly to the removal budget.
    \item We propose a confidence-adaptive audit that identifies hard-to-detect clean-label attacks through candidate clustering and uses causal removal tests to verify suspected backdoors, failing safely when evidence is not enough.
\end{itemize}
To validate our proposal, we performed a thorough experimental validation over four benchmarking datasets and four common model architectures. The membership analysis confirms that our auditing lens captures a genuine internalization signal, while the backdoor identification achieves over $90\%$ Top-1 on most settings while remaining the most consistent method across the grid.

\section{Background}
\label{sec:background}
This section is dedicated to presenting the core concepts relevant to this work, including low-rank adaptation methods for parameter-efficient fine-tuning, membership inference attacks that threaten training data privacy, and backdoor attacks that compromise model integrity.

\subsection{Low-Rank Adapters}
\label{subsec:lora}

State-of-the-art deep models are typically pre-trained on large corpora and adapted to downstream tasks through fine-tuning. Fine-tuning all parameters is computationally costly for large models, which motivates methods that adapt only a small subset of parameters. Parameter-efficient fine-tuning (PEFT) methods address this challenge by updating only a small subset of parameters while keeping the pretrained weights fixed. Low-Rank Adaptation (LoRA) is a representative PEFT method that constrains weight updates to a low-rank subspace~\cite{hu2022lora}: instead of directly updating a weight matrix $W \in \mathbb{R}^{d \times k}$, it introduces a low-rank decomposition of the weight update $\Delta W = BA$, with $B \in \mathbb{R}^{d \times r}$, $A \in \mathbb{R}^{r \times k}$ and $r \ll \min(d, k)$, so that the adapted layer computes $h = Wx + \alpha BAx$ with scaling factor $\alpha = 1/r$. By reducing the number of trainable parameters, LoRA enables efficient fine-tuning while preserving model performance. The modular nature of adapter parameters also introduces new security considerations, as adapters can be independently trained, shared, or modified~\cite{fu2024loft, mia2025fedshield}. Given the demonstrated effectiveness of steering pretrained large models toward desired behaviors by updating only a small number of parameters, we were motivated to investigate whether such adapters can efficiently encode essential information from the pretrained model itself.

\subsection{Membership Inference Attacks}
\label{subsec:mia}

Membership inference attacks (MIAs)~\cite{niu2024survey} aim to determine whether a particular data sample was included in the training dataset of a machine learning model, thereby violating data privacy~\cite{shokri2017membership}. Let $\mathcal{D}_{\text{pre}}$ denote the training dataset used to learn a model $f_\theta$; given a sample $x$ and the model output $f_\theta(x)$, the attacker infers whether $x\in\mathcal{D}_{\text{pre}}$, formalized as a function $\mathcal{A}:(x, f_\theta(x)) \to \{0,1\}$ that maximizes the inference accuracy $\Pr[\mathcal{A}(x, f_\theta(x)) = m(x)]$ for the membership variable $m(x)$. A common class of MIAs exploits the discrepancy between training and non-training samples, often quantified using the model loss $\ell(x) = \mathcal{L}(f_\theta(x), y)$: samples that obtain lower loss values are more likely to be inferred as members~\cite{yeom2018privacy}, while more advanced attacks leverage confidence scores, gradients, or shadow models~\cite{carlini2022membership}. In this work membership inference is not a target contribution but the setting with ground-truth internalization, which we use to validate that our low-rank lens reads what a model has absorbed before applying it to backdoors.

\subsection{Backdoor Attacks}
\label{subsec:backdoor}

Backdoor attacks~\cite{bai2024backdoor} are training-time attacks that aim to embed hidden malicious behavior into a model while preserving its normal performance on clean inputs: the model behaves normally on benign, unmodified samples but reliably exhibits attacker-chosen behavior when specific triggers are present. Backdoors can be introduced through several attack vectors; in this work we focus on data poisoning–based attacks, which are the most commonly studied and widely applicable. Given a clean training dataset $\mathcal{D} = \{(x_i, y_i)\}_{i=1}^{N}$, the attacker selects a trigger $t$ and applies a trigger injection function $\phi(\cdot)$ to obtain a poisoned input $\hat{x} = \phi(x, t)$, poisoning a fraction $\epsilon = M/N$ ($M \ll N$) of the data. Depending on whether the attacker modifies the label, attacks are categorized as \emph{dirty-label} (the original label is replaced with an attacker-chosen target label $\hat{y}$) or \emph{clean-label} (the ground-truth label remains unchanged). Triggers can further be static (sample-agnostic) or dynamic (sample-specific), and source-specific or source-agnostic~\cite{gu2019badnets, chen2017targeted, nguyenwanet, turner2019label, hong2022handcrafted}. The effectiveness of a backdoor attack is typically evaluated using the Attack Success Rate (ASR), the fraction of triggered test inputs classified as the target label, while clean accuracy measures the model's performance on unmodified samples, ensuring the backdoor does not significantly degrade benign behavior.

\section{Methodology}
\label{sec:method}
We first define the auditing setting (Section~\ref{sec:threatModel}), then develop our lens along the progression of our argument: we stablish it on membership, where ground truth provides direct validation (Section~\ref{sec:inferenceMode}); apply it to detect a backdoor's malicious internalization and identify its target (Section~\ref{sec:backdoorMode}); and, finally, act on it by erasing the backdoor in the same low-rank subspace, with detection confidence sizing the repair (Section~\ref{sec:removalMode}).

\subsection{Threat Model}
\label{sec:threatModel}
We consider a defensive post-training auditor who receives a pre-trained model $f_\theta$ and must verify the presence of existing backdoors in the model. 
The pre-trained model $f_\theta$ is obtained by training on an unknown dataset $D_{\mathrm{pre}}$, which may contain poisoned samples injected by an adversary. The existence of a backdoor in the model is therefore not assured. Consequently, the auditor may regard the model as benign and elect not to perform any backdoor removal or model sanitization procedures.
The auditor does not have access to the original training data $D_{\mathrm{pre}}$, so holds no subset of the poisoned  $D_p \in D_{\mathrm{pre}}$, has no knowledge on the nature of the trigger, and cannot activate an existing backdoor at inference. Consistent with deployment practice, the only permitted intervention is parameter-efficient fine-tuning, while full retraining and unrestricted gradient access are disallowed.
The auditor holds a small, held-out clean reference set $D_{\mathrm{clean}}$ of in-distribution samples, containing neither triggered nor poisoned inputs and disjoint from $D_{\mathrm{pre}}$. This is our only data assumption, which matches the access granted to competing defenses; we never assume access to poisoned data, the training set, or trigger knowledge. Under this setting, the audit has two objectives: (i) identify a backdoor target class $c_t\in\mathcal{Y}$; (ii) given the target(s), sanitize $f_\theta$ so ASR collapses while Clean Data Accuracy (CDA) is preserved.

\subsection{LoRAcle}
The LoRAcle auditing strategy is designed to leverage Low-Rank Adaptation (LoRA) to efficiently adapt a pretrained model $f_\theta$ to a new dataset. Our underlying hypothesis, grounded in this property, is that a model that has already internalized specific features, whether benign (clean) or associated with triggers, will exhibit behavior distinct from that observed on previously unseen features.
By internalization, we mean the strong association that the weights $W$ have learned with either the features of a memorized training example or an attacker's trigger-to-target shortcut. This association provides the signal that our lens exploits to identify the backdoor target class $c_t\in\mathcal{Y}$.
Fine-tuning a pretrained model toward content it has already internalized moves its parameters a short, aligned distance, whereas adapting toward novel content demands a larger, less aligned update. The geometry of the adaptation $\Delta W$ (its cosine alignment with $W$) therefore acts as an oracle for prior internalization. 

\begin{figure}[]
\centering
\includegraphics[width=0.8\linewidth]{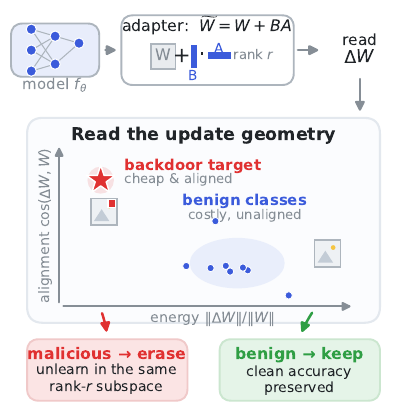}
\caption{\textbf{\textsc{LoRAcle}.}}
\label{fig:teaser}
\end{figure}

\subsubsection{Internalization Lens}
\label{sec:inferenceMode}

Borrowing some intuition from membership inference attacks, we infer the internalization of a set of features, verifying the low-rank geometry of a LoRA as an auditing tool. Given a batch $B$, we fine-tune a LoRA adapter on $B$ using its labels and measure the geometry of the resulting parameter update (this method will be then used in Section~\ref{sec:backdoorMode}). Specifically, we consider its relative energy, $E=\|\Delta W\|/\|W\|$, and the cosine alignment between $\Delta W$ and the frozen weights $W$. If the model has already internalized the information in $B$, adapting to it requires only a small update that is strongly aligned with the existing weights. In contrast, adapting to an unfamiliar batch requires a larger and less aligned update.

To avoid using a separate reference or calibration set, we compare the update obtained with the correct labels against a self-null obtained by randomly shuffling the labels:
 
\begin{equation}
    r=\|\Delta W_{\mathrm{shuf}}\|/(\|\Delta W_{\mathrm{shuf}}\|+\|\Delta W_{\mathrm{corr}}\|),
\end{equation}
which places a non-member at an absolute $0.5$ with no calibration set. 

It is important to stress that the solution described in this section is intended as a diagnostic instrument, rather than as a novel membership attack. For models that memorize, a simple behavioral test already separates members from non-members almost perfectly. Therefore, the relevant value of our lens lies in settings where observable behavior and internalization diverge (Section \ref{sec:analysis}, \ref{sec:results}).

\subsubsection{Backdoor Detection}
\label{sec:backdoorMode}
Building on this view of the lens as a probe of internalization, we now apply the same intuition to backdoor detection, where internalization takes a malicious form: the poisoned network was driven to absorb a shortcut to a target $c_t$, which can be seen as a ``membership'' planted by the attacker. Therefore, when $c=c_t$, fitting a diagnostic adapter to a candidate $c$ should encounter the same reduced resistance that our internalization hypothesis predicts for genuine members. We read this through two complementary views that share one readout.

To infer the membership of a backdoor given the pre-trained model $f_\theta$, for each candidate $c$ we fit a rank-$r$ adapter and read $\Delta W_c$ at a fixed architecture-appropriate layer $\ell$ attached to the first fully connected layer after the feature extractor. 
To extract the geometry of the model, we train a LoRa adapter using two probes: the clean probe $p_{clean}$ that fits on the auditor's clean class-$c$ samples and the data-free (smooth) probe $p_{smooth}$ that fits on a synthetic proxy batch generated from the model itself,
\begin{equation}
   x_i^{(t+1)}=x_i^{(t)}-\eta\nabla_{x_i}\mathcal{L}_{\mathrm{proxy}}(f_\theta(x_i^{(t)}),c)+\sigma\epsilon_t, 
\end{equation}
with $\mathcal{L}_{\mathrm{proxy}}$ combining cross-entropy, contrastive and smoothness terms.

The two probes are complementary by design, differing precisely in their data assumptions, and each earns its place. The smooth probe is data-free and label-free, and it fabricates its own class-$c$ inputs from the model, so the audit degrades gracefully to a regime with no clean data. Thus, being synthesized independently of the reference set, contributes a second, uncorrelated view of the same geometry. The clean probe instead consumes the small labeled reference set $D_{\mathrm{clean}}$, which groups in-distribution samples by their class, to provide a sharper, higher-fidelity nomination. Keeping both is therefore not redundancy but coverage on two fronts: the data-free probe secures the no-clean-data regime, the clean probe the high-fidelity one, and, critically, their two nominations supply a confidence signal that a single probe cannot, as we exploit next. We stress that ``label-free'' qualifies the target inference, with neither probe having prior knowledge of the backdoor target and neither the trigger nor any triggered data ever being seen.

To detect the target class, from $\Delta W_c$ at $\ell$ we extract the relative energy, cosine alignment with the frozen weights, and spectral stable rank,
\begin{equation}
E_c=\frac{\|\Delta W_c\|_2}{\|W\|_2},\quad
C_c=\frac{\langle W,\Delta W_c\rangle}{\|W\|_2\|\Delta W_c\|_2},
\end{equation}
each robustly $z$-normalized across classes. For $s\in\{C,E\}$,
  \begin{equation}
  z_s(c)=0.6745(s_c-\mathrm{median}(s))/\mathrm{MAD}(s).
  \end{equation}

Intuitively, if a backdoor is present for a given target class, we expect high alignment and low relative energy, as a backdoor effectively introduces a shortcut that causes the model to be highly confident in its prediction.
Hence, we combine high alignment with low relative energy, as the planted shortcut is a direction the pretrained weights already support, thus adapting towards it requires only a small and well-aligned update. This is the entire rule,
\begin{equation}
P_c = z_C(c) - z_E(c),
\label{eq:fusion}
\end{equation}
with no tuned thresholds. 
For each of the two probes $p_{clean}$ and $p_{smooth}$, we rank the classes to detect the most probable target. The nomination is done by $\hat c=\arg\max_c P_c$, aggregated over multiple trials to ensure rank stability. The two terms are not redundant, and neither suffices alone, as we will prove in Section \ref{sec:ablation}.
The two perspectives lead to nominations $\hat c_{\mathrm{clean}},\hat c_{\mathrm{smooth}}$
whose agreement is a label-free confidence signal (\rev{Table~\ref{tab:overview}}). In practice, agreement commits to a single class; disagreement widens to the smallest set spanning the two views. This uncertainty-aware set is what governs repair.

\begin{table}[t]\centering\scriptsize
\caption{The two-probe consensus as a label-free confidence gate ($89$ dirty-label victims at poison rate $0.2$; four datasets $\times$ four architectures $\times$ BadNets/Blend/WaNet): agreement commits to a single class ($K{=}1$), disagreement widens the budget.}
\label{tab:overview}
\begin{tabular}{l ccc}
\toprule
Consensus & \% of models & budget $K$ & target recall (\%)\\
\midrule
Agree $\to$ commit & $69.7$ & $1$ & $98.4$\\
Disagree $\to$ widen & $30.3$ & $\sim\!2$ & $63.0$\\
\midrule
Adaptive (overall) & $100$ & $1.30$ & $87.6$\\
\bottomrule
\end{tabular}
\end{table}

\subsubsection{Backdoor Removal}
\label{sec:removalMode}

The same property used to identify a backdoor target can also be used to erase the trigger knowledge from the model. We sanitize within the adapter subspace, never retraining the
backbone, and, distinctively, let the detection confidence set how much to erase. The unlearning step itself is deliberately unremarkable, fine-tuning a low-rank adapter to forget a class is routine; the contribution is the \emph{coupling}. The same confidence signal that identifies the target also sizes and gates the repair, so detection and removal are one mechanism rather than two separately-tuned stages.

A backdoor is a mapping the model internalized, a trigger, wherever it appears, is routed to the target
class $c$. To erase it, we first recover a synthetic trigger that activates the shortcut, then
suppress that activation in the adapter subspace while leaving clean behavior intact.

The first step is recovery of the trigger. Given the detected target $c$ and the clean set $D_{\mathrm{clean}}$,
we reverse-engineer a mask $m\in[0,1]^{H\times W}$ and pattern $p$ that make the clean inputs collapse onto $c$,
\begin{equation}
\min_{m,p}\ \ \mathbb{E}_{x\sim D_{\mathrm{clean}}}\big[\mathcal{L}_{\mathrm{CE}}\big(f_\theta(x\odot(1-m)+p\odot m),\,c\big)\big] \;+\; \lambda\,\|m\|_1 ,
\label{eq:recon}
\end{equation}
where the $\ell_1$ penalty drives the mask to the smallest region that still flips the prediction;
we keep the smallest-mask solution over random restarts. A compact recovered mask means a genuine
patch/blend-type trigger has been localized; a large, diffuse mask signals that the true trigger is
input-wide (e.g.\ a warping field) and cannot be captured this way.

Once the trigger is recovered, it can be unlearned using our lightweight strategy. Since the trigger can result either as a compact set of features or as a diffuse, spurious pattern, we need to address the unlearning using two different strategies.
First, a surgical unlearn for compact triggers. We build an unlearning set by stamping the recovered trigger onto clean inputs, writing $\tilde{x}=x\odot(1-m)+p\odot m$ for the stamped input, but keeping their true labels $y$. Then, we train only the rank-$r$ adapter $\Delta W=BA$ (the backbone $W$ frozen), interleaved with clean data to preserve accuracy,
\begin{equation}
\min_{A,B}\ \ \mathbb{E}_{(x,y)}\Big[
  \mathcal{L}_{\mathrm{CE}}\big(f_{W+\Delta W}(\tilde{x}),\,y\big)
  + \mathcal{L}_{\mathrm{CE}}\big(f_{W+\Delta W}(x),\,y\big)\Big].
\label{eq:unlearn}
\end{equation}
The adapter learns to route triggered inputs back to their correct class, removing the shortcut inside
the same low-rank subspace that exposed it, a surgical edit that alters a small amount of parameters rather than a
full-model retrain.
If we obtain a diffuse trigger instead, clean fine-tuning is performed. When Equation~\ref{eq:recon} returns a large, spurious
mask, there is no compact operator to unlearn; instead, we briefly fine-tune on held-out clean data.
Input-wide triggers have no compact operator to excise, but because they perturb the entire image they overlap heavily with features that clean inputs already exercise; a short pass over unseen clean data therefore suppresses them without a reconstructed trigger, as our removal results confirm (Section~\ref{sec:results}). The one adversarial exception is a physics-stealth full-image trigger, which becomes metastable under removal rather than converging (Section~\ref{sec:extended}), a boundary we report explicitly. The contribution lies not in the fallback itself, but in determining, without triggered data, when to invoke it, and performing the localizable case as a surgical adapter edit rather than a whole-model pass.
The two arms are not chosen blindly, as the mask size from Eq.~\ref{eq:recon} routes between them without triggered data. A compact mask certifies that reconstruction has localized a genuine operator and routes to the surgical edit; only a large, diffuse mask, evidence that reconstruction has failed, defers to the clean fine-tune. This makes surgical edit the primary removal strategy, as the reconstruction is compact and removable in most evaluated settings (${\sim}71\%$), while clean fine-tuning is reserved for the input-wide remainder (Section~\ref{sec:results}).

Since our detection method returns a ranking of possible targeted classes, the number of classes to sanitize is determined by the consensus: on agreement we unlearn one class ($K{=}1$); on disagreement we unlearn the minimum cluster
\begin{equation}
S = \mathrm{elbow}(P^{\mathrm{clean}}) \cup \mathrm{elbow}(P^{\mathrm{smooth}}),
\label{eq:cluster}
\end{equation}
where $\mathrm{elbow}(P)$ cuts the sorted scores at their largest gap (capped at a small $K_{\max}$). Since sanitizing additional classes may reduce clean accuracy and agreement signals a reliable nomination, this spends the smaller budget for the confident majority and the wider budget only when the audit is uncertain, yielding a Pareto-efficient alternative to a fixed top-$K$ removal budget with no tunable threshold.

When the cluster $S$ holds more than one candidate, we make the final selection causal rather than score-based. We unlearn each candidate in $S$ in turn and keep the one whose removal actually collapses the attack. Indeed, only editing the true target's shortcut destroys the backdoor, whereas editing a decoy leaves ASR intact. This causal adjudicator costs $|S|$ lightweight rank-$8$ edits and is what allows the pipeline to recover a target that the probe scores rank below threshold, converting an uncertain audit into a confirmed one rather than a confident guess. This is the mechanism that preserves the method’s effectiveness precisely where point identification is weakest (Section~\ref{sec:narcissus})..

\textbf{Scope and intent}
We do not propose a new state-of-the-art detector or eraser. The contribution is the principle, a lens on internalization grounded in membership inference, and the efficient detect-and-erase pipeline it enables. One low-rank mechanism first shows that it can read what a model absorbed, then uses that signal to find and remove a backdoor's malicious internalization, with detection confidence sizing the repair. Our comparisons to recent defenses (Section~\ref{sec:results}) show that the lens remains competitive while being more efficient.

\section{Validating the Lens on Ground Truth}
\label{sec:analysis}
In this section, we validate our approach as lens to find known ``members'' in a pre-trained model before focusing on backdoors. 
We must establish that its geometry tracks which features a model has internalized. Membership inference is the natural test bench, even if we are not proposing a membership inference attack, it is a perfect test bench to validate the pros and the limits of the approach. On models that memorize, a trivial behavioral test (per-batch loss or predictive entropy) already separates members from non-members almost perfectly (Table~\ref{tab:membership}), and we do not contest that. The reason is instructive: for membership, behavior and internalization coincide, a
memorized sample is both confidently classified and geometrically encoded, so the cheapest behavioral
proxy wins. We use this setting only to confirm that the low-rank geometry reads the internalization axis
too, before turning to backdoors, where behavior and internalization diverge and the behavioral
proxy fails.

\subsection{The Geometry Tracks Internalization}
\label{sec:memresults}
The lens turns the geometry of Section~\ref{sec:inferenceMode} into a score: a member batch induces a
small, well-aligned, stable low-rank update, a non-member batch a large, weakly aligned, chaotic one. The results in Table~\ref{tab:membership} show that the signal is real. Across the memorizing cells it ranks members above non-members well above chance (Table~\ref{tab:membership}: LoRA AUROC $0.94$-$0.98$ on CIFAR-10), and it tracks the graded
member-fraction of a batch monotonically (Spearman $\approx 0.96$ on CIFAR-10-ResNet-18). It also fails
where it should on MNIST, whose models barely memorize, the LoRA signal drops toward chance
($0.69$), and on the 100-class CIFAR-100 it weakens to $0.57$-$0.84$, the same architectural/scale axis
on which the backdoor lens softens in Section~\ref{sec:results}, evidence that it reads a genuine
internalization gap rather than an artifact. What it does not do is beat the behavioral baselines,
loss and entropy sit at $0.97$-$1.00$ on every memorizing cell, because membership is behaviorally
saturated. That is the expected and honest outcome, and it is precisely why membership is a validation
here and not a result.

\begin{table}[t]\centering\scriptsize
\caption{Membership inference validation, ranking AUROC (member-fraction $>0$).}
\label{tab:membership}
\begin{tabular}{ll ccc}
\toprule
 & & \multicolumn{2}{c}{behavioral (label-free)} & LoRA lens\\
\cmidrule(lr){3-4}\cmidrule(lr){5-5}
Dataset & Model & loss & entropy & (ours)\\
\midrule
CIFAR10 & resnet18 & 1.00 & 1.00 & 0.98\\
CIFAR10 & vgg19 & 0.97 & 0.99 & 0.96\\
CIFAR10 & densenet & 0.97 & 1.00 & 0.94\\
CIFAR10 & vit & 1.00 & 1.00 & 0.95\\
CIFAR100 & resnet18 & 1.00 & 1.00 & 0.57\\
CIFAR100 & vgg19 & 1.00 & 1.00 & 0.84\\
CIFAR100 & densenet & 1.00 & 1.00 & 0.82\\
\midrule
MNIST$^{\dagger}$ & resnet18 & 0.80 & 0.88 & 0.69\\
\bottomrule\end{tabular}
\\[2pt]{\scriptsize $^{\dagger}$ Negative control: MNIST models barely memorize, so the internalization
gap, and the LoRA signal ($0.69$), shrinks toward chance, as expected. Even the behavioral baseline
weakens here ($0.80$/$0.88$), confirming the effect is memorization, not an artifact.}
\end{table}

\subsection{What the Lens Reads: Familiarity, Not Memorization}
\label{sec:continuum}
\begin{figure}[t]\centering
\includegraphics[width=0.82\linewidth]{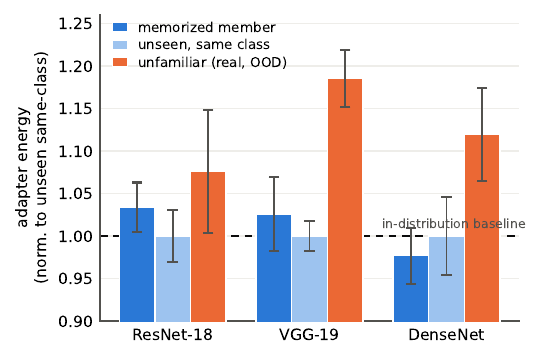}
\caption{What the geometry measures. Energy to adapt toward a class, normalized per architecture to the
unseen-same-class baseline.}
\label{fig:continuum}
\end{figure}

In Section~\ref{sec:memresults}, we validated that LoRa can be used to track the internalization of features in a pre-trained model but a question remains: does the lens read memorization (this exact sample was in training) or
representational familiarity (the model has structure for this kind of content)? An unseen but
in-distribution image, a held-out airplane the model classifies perfectly, would look like a member
under the second reading and a non-member under the first. We answer it directly. On a clean CIFAR-10
model we fit the diagnostic adapter toward a class and measure its energy for four batch types: memorized
training images of that class, unseen test images of the same class, unfamiliar real images
(CIFAR-100 categories the model never learned), and Gaussian noise (Fig.~\ref{fig:continuum}).

Two facts hold across three architectures. First, memorized and unseen same-class batches are indistinguishable (within $1\sigma$, inconsistent sign): storing the specific instances adds nothing to the geometry. Second, unfamiliar real content costs more to encode than in-distribution content. The lens therefore reads representational internalization, whether the weights carry structure for the content, not instance memorization. This is why ``internalization'' and not ``memorization'' is the right word; it is why membership (an instance-level, behavioral property) is better read by loss than by our geometry; and it is why a backdoor, a learned representational shortcut from trigger to target, is the kind of feature the lens does read (Section~\ref{sec:results}).

Two properties of the lens are worth carrying forward. First, the lens is reference-free: the ratio score's shuffled-label self-null gives an absolute decision anchor (a non-member sits at $0.5$ by construction), so it needs no calibration set; we stress this is a convenience, not an advantage over the baselines, which here are stronger. Second, and more important, the lens reads a different axis than behavior: on membership the two axes coincide, so the cheap behavioral proxy wins, but they come apart on backdoors, which can be behaviorally loud yet geometrically faint, or behaviorally quiet yet geometrically obvious (Section~\ref{sec:results} shows both). That divergence, not a better membership number, is what the lens is for. We therefore treat membership as the sanity check it is, and make no claim of a membership attack.

\section{Evaluation}
\label{sec:results}
\subsection{Experimental Setup}
We evaluate on MNIST, CIFAR-10, GTSRB and CIFAR-100 across ResNet-18, VGG-19, DenseNet and ViT, against
BadNets, Blend and WaNet triggers at poison rate $0.2$ (full grid and hyperparameters in Appendix~\ref{app:impl}). These triggers are chosen to span the trigger natures a weight-space auditor must detect: BadNets is a small, localized, high-frequency patch (a compact, well-encoded shortcut), Blend a diffuse full-image blended pattern, and WaNet an imperceptible warping that leaves no additive pixel-space signature and so stresses reconstruction-based detectors in particular. We further add the clean-label Narcissus attack (Section~\ref{sec:narcissus}), whose full-image perturbation keeps poisoned images correctly labelled and thus leaves no dirty-label class shortcut to read, as the most adversarial case for a weight-space reader. Poisoning rate analysis over lower rates $\{0.01,0.05,0.1\}$ is in Section~\ref{sec:dose}. Detection accuracy is reported as a Top-1/Top-3 rate if the taget class is the first candidate or is between the $3$ most probable classes.
Removal reports residual ASR (lower better) and clean accuracy (CDA, higher better) under the guarded best-epoch rule, with a clean set of $5\%$ of the training data granted to every method~\cite{lineural,wu2021adversarial,zengadversarial,liu2018fine}. Baselines are chosen to represent the distinct families of post-training defense, detection (Neural Cleanse, K-Arm, MM-BD) and removal (Fine-Pruning, ANP, NAD, I-BAU), so the comparison verifies our method against the mechanism each family embodies rather than a single implementation (method details in Appendix~\ref{app:impl}, citations in Section~\ref{sec:related}).

\subsection{Backdoor Detection}
\label{sec:detection}
Within this section, we start the evaluation campaign of LoRAcle against backdoors, and in this particular one we will start by measuring its performance in detecting backdoor target classes.
Results are reported in Table~\ref{tab:detection},
Among methods in our own threat model, LoRAcle leads on Top-1, well clear of Neural Cleanse, whose pixel-space inversion is weak on WaNet, and K-Arm. Against MM-BD, the strongest recent competitor (which uses no data), the two are close on aggregate. We deliberately do not compare against detectors like ScaleUp, Spectral Signatures, Activation Clustering, as they require the poisoned training set, outside our threat model, and, under dirty-label poisoning at rate $0.2$, the target class holds $\approx 28\%$ of the training data, so a trivial control that merely counts
labels identifies the target for free. Those methods do not beat that control; their data access, not their
algorithm, does the work, so a target-identification comparison against them is not meaningful.

\begin{table}[t]\centering\scriptsize
\caption{Target identification at poison rate $0.2$.}
\label{tab:detection}
\resizebox{\columnwidth}{!}{%
\begin{tabular}{l cc l}
\toprule
Method & Top-1 (\%) & Top-3 (\%) & Data required \\
\midrule
\textbf{\textsc{LoRAcle} (ours)} & \textbf{86.7} & \textbf{94.3} & clean + weights \\
MM-BD & 80.0 & 91.1 & weights \\
K-Arm & 79.6 & 88.2 & clean + weights \\
Neural Cleanse & 35.1 & 43.2 & clean + weights \\
\bottomrule
\end{tabular}}
\end{table}

\begin{figure}[t]\centering
\includegraphics[width=\linewidth]{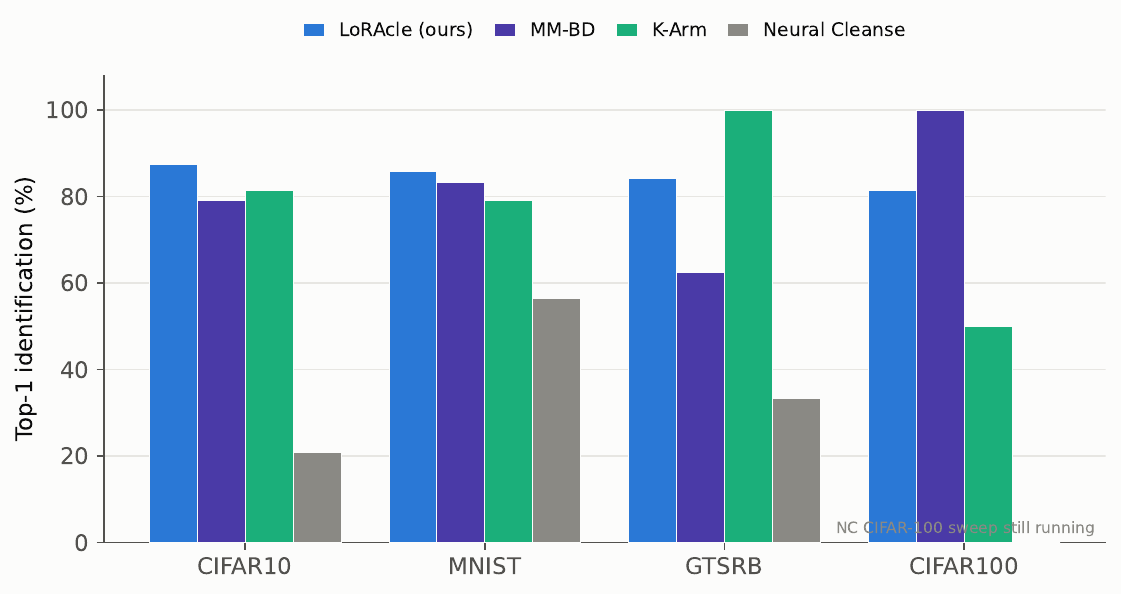}
\caption{Per-dataset Top-1 identification at poison rate $0.2$.}
\label{fig:detbar}
\end{figure}

The granular view in Table~\ref{tab:det-granular} and Figure~\ref{fig:detbar} is the more informative one. The baselines each have a collapse cell (MM-BD drops to $0$-$50\%$ on GTSRB; K-Arm to $50\%$ on CIFAR-100), whereas LoRAcle stays in the $67$-$100\%$ band on every architecture$\times$dataset cell. Its lead comes from never collapsing, not from dominating any one benchmark.

\begin{table}[t]\centering\scriptsize
\caption{Granular Top-1 target identification per architecture$\times$dataset at poison rate $0.2$.}
\label{tab:det-granular}
\resizebox{\columnwidth}{!}{%
\begin{tabular}{ll cccc}
\toprule
Arch & Dataset & \textbf{LoRAcle} & MM-BD & K-Arm & NC\\
\midrule
\multirow{4}{*}{ResNet-18} & CIFAR-10 & \textbf{96.7$\pm$3.3} & 100.0$\pm$0.0 & 88.9$\pm$10.5 & 16.7$\pm$15.2\\
 & MNIST & \textbf{93.3$\pm$4.6} & 83.3$\pm$15.2 & 83.3$\pm$15.2 & 66.7$\pm$19.2\\
 & GTSRB & \textbf{100.0$\pm$0.0} & 50.0$\pm$20.4 & 100.0$\pm$0.0 & 33.3$\pm$19.2\\
 & CIFAR-100 & \textbf{93.3$\pm$6.4} & 100.0$\pm$0.0 & 50.0$\pm$20.4 & --\\
\midrule
\multirow{4}{*}{VGG-19} & CIFAR-10 & \textbf{93.3$\pm$4.6} & 33.3$\pm$19.2 & 83.3$\pm$15.2 & 16.7$\pm$15.2\\
 & MNIST & \textbf{83.3$\pm$6.8} & 83.3$\pm$15.2 & 83.3$\pm$15.2 & 66.7$\pm$19.2\\
 & GTSRB & \textbf{100.0$\pm$0.0} & 0.0$\pm$0.0 & 100.0$\pm$0.0 & 33.3$\pm$19.2\\
 & CIFAR-100 & \textbf{100.0$\pm$0.0} & 100.0$\pm$0.0 & 50.0$\pm$20.4 & --\\
\midrule
\multirow{4}{*}{DenseNet} & CIFAR-10 & \textbf{90.0$\pm$5.5} & 100.0$\pm$0.0 & 66.7$\pm$19.2 & 16.7$\pm$15.2\\
 & MNIST & \textbf{80.0$\pm$7.3} & 100.0$\pm$0.0 & 83.3$\pm$15.2 & 33.3$\pm$19.2\\
 & GTSRB & \textbf{73.3$\pm$8.1} & 100.0$\pm$0.0 & 100.0$\pm$0.0 & 0.0$\pm$0.0\\
 & CIFAR-100 & \textbf{66.7$\pm$12.2} & 100.0$\pm$0.0 & 50.0$\pm$20.4 & 0.0$\pm$0.0\\
\midrule
\multirow{3}{*}{ViT} & CIFAR-10 & \textbf{76.7$\pm$7.7} & 83.3$\pm$15.2 & 83.3$\pm$15.2 & 33.3$\pm$19.2\\
 & MNIST & \textbf{86.7$\pm$6.2} & 66.7$\pm$19.2 & 66.7$\pm$19.2 & 60.0$\pm$21.9\\
 & GTSRB & \textbf{66.7$\pm$8.6} & 100.0$\pm$0.0 & 100.0$\pm$0.0 & 66.7$\pm$19.2\\
\midrule
\multicolumn{2}{l}{\textbf{Overall}} & \textbf{86.7$\pm$1.7} & 80.0$\pm$4.2 & 79.6$\pm$4.2 & 35.1$\pm$5.1\\
\bottomrule\end{tabular}}
\end{table}

\textbf{Detection efficiency}
On the detection side the memory gap is architecture-dependent: reconstruction-based detection must backpropagate to the input and store all activations, costing $8.2\times$ more peak GPU memory than our probe on DenseNet, while the two are comparable on ResNet-18 and ViT (Appendix~\ref{app:effdet}).

\subsection{What the Lens Actually Reads: Fusion Ablation}
\label{sec:ablation}
The identification rule (Equation~\ref{eq:fusion}) is a difference of two normalized quantities, so it is worth asking whether either term is contributing alone to the detection, and whether a richer rule would do better. We answer both by re-scoring the same $405$ probed models under every variant, reported as Top-1/Top-3 (\%): alignment alone reaches $67.2$/$79.0$, energy alone $73.3$/$83.4$, and spectral alone $18.8$/$27.4$; the deployed $z_C-z_E$ reaches $\mathbf{86.7}$/$\mathbf{94.3}$, adding the spectral term ungated drops it to $73.6$/$87.8$, and a gated spectral term reaches $86.2$/$94.6$.

Two things follow. First, the mechanism is joint; alignment alone reaches $67.2$ and energy alone $73.3$, but their difference reaches $86.7$, a target is not merely well-aligned or merely cheap to adapt towards; it is both at once, and only the contrast isolates it. This is the sharpest evidence that the lens reads a genuine internalization signature rather than a single convenient statistic. Second, a more elaborate rule does not help. Adding the spectral term ungated is actively harmful ($-13.1$), and admitting it through a tuned gate recovers to $86.2$, still below the plain difference. We tested the gate's two thresholds over a $6\times5$ grid, and the score increases monotonically as the gate is made stricter, peaking exactly where it stops firing altogether. We therefore report the simpler rule: it is more accurate, it raises the worst cell in Table~\ref{tab:det-granular} from $60.0$ to $66.7$, and it leaves the method with no free parameters to tune.

\subsection{The Signal Reads Internalization, Not Optimization Difficulty}
\label{sec:rankabl}

\begin{figure}[t]\centering
\includegraphics[width=0.78\linewidth]{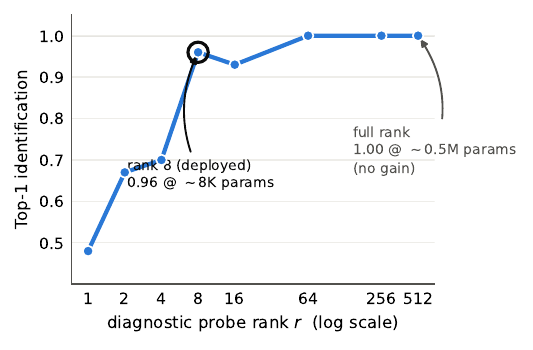}
\caption{Identification vs.\ diagnostic-probe rank.}
\label{fig:rank}
\end{figure}

In this section, we want to verify if our identification signal is an artifact of the low-rank probe, whether constraining the adapter is what produces the alignment we read. We test this directly by testing the diagnostic adapter's rank from $1$ to full ($512$) on $27$ CIFAR-10 backdoored models per rank and re-measuring Top-1 identification (Figure.~\ref{fig:rank}). The signal strengthens with rank rising from $0.48$ at rank $1$ to $0.96$ at rank $8$ and saturating at $1.00$ by rank $64$, holding there at full rank. Two conclusions follow. First, the mechanism is not the low-rank constraint; a full-rank adapter, the easiest optimization, produces the best target separation, so the signal cannot be an artifact of low-rank fitting being hard. It is the alignment of adapting toward internalized content, present at every rank. Second, this justifies rank-$8$ empirically, it sits at $0.96$, within a few points of the full-rank ceiling, for roughly two orders of magnitude fewer parameters ($\sim\!10^4$ vs $\sim\!10^6$). The low rank buys the parameter and memory efficiency (Figure~\ref{fig:effrm}) at almost no cost in signal. More broadly, the signal reads on the adaptation direction rather than the rank constraint, so any parameter-efficient probe that exposes an update geometry is a candidate reader.

The rank testing already shows that we merely read how hard the fit is, because the easiest optimization (full rank) yields the best separation, the opposite of a difficulty artifact. Three further controls converge on the same reading. First, the familiarity experiment (Section~\ref{sec:continuum}) shows the energy tracks representational familiarity, not instance memorization or raw fit cost. Second, the fusion ablation (Section~\ref{sec:ablation}) shows the signal is the joint low-energy-and-aligned contrast, alignment alone reaches only $67.2$, energy alone $73.3$, their difference $86.7$, so it is no single landscape statistic. Third, an untrained-model control (Appendix~\ref{app:untrained-control}). On a random-init model of the same architecture, an identical optimization problem with nothing internalized, the familiar-versus-unfamiliar energy ordering that separates content at AUROC $0.83$--$0.99$ on trained models collapses to chance ($0.21$--$0.25$), with all batch types inside measurement noise. Optimization geometry is a correlate the lens is robust to, not the quantity it measures.

\subsection{Backdoor Removal}
\begin{figure}[t]\centering
\includegraphics[width=0.82\linewidth]{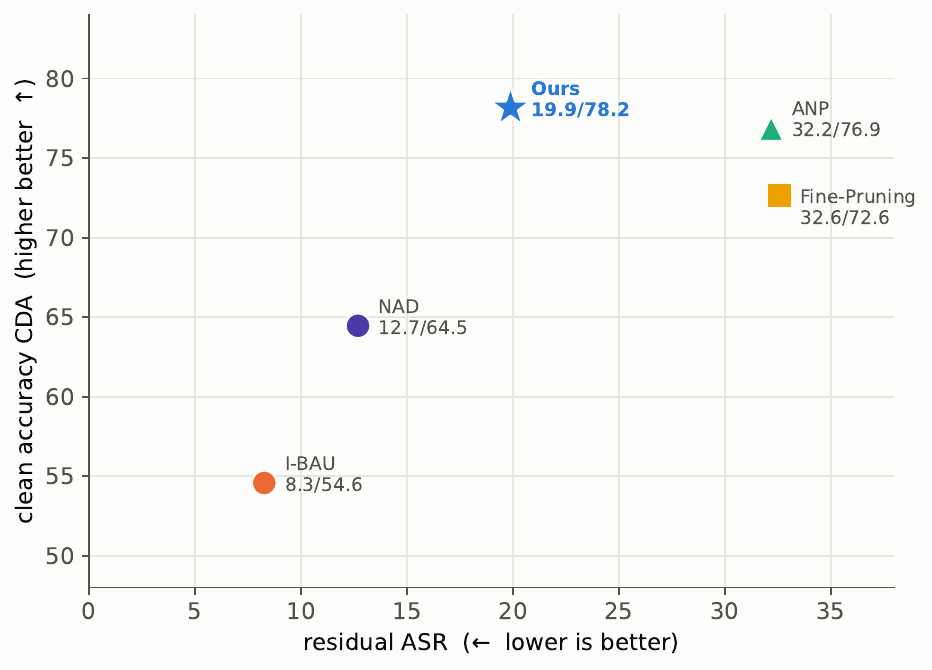}
\caption{Removal trade-off.}
\label{fig:rmpareto}
\end{figure}

\begin{figure}[t]\centering
\includegraphics[width=0.72\linewidth]{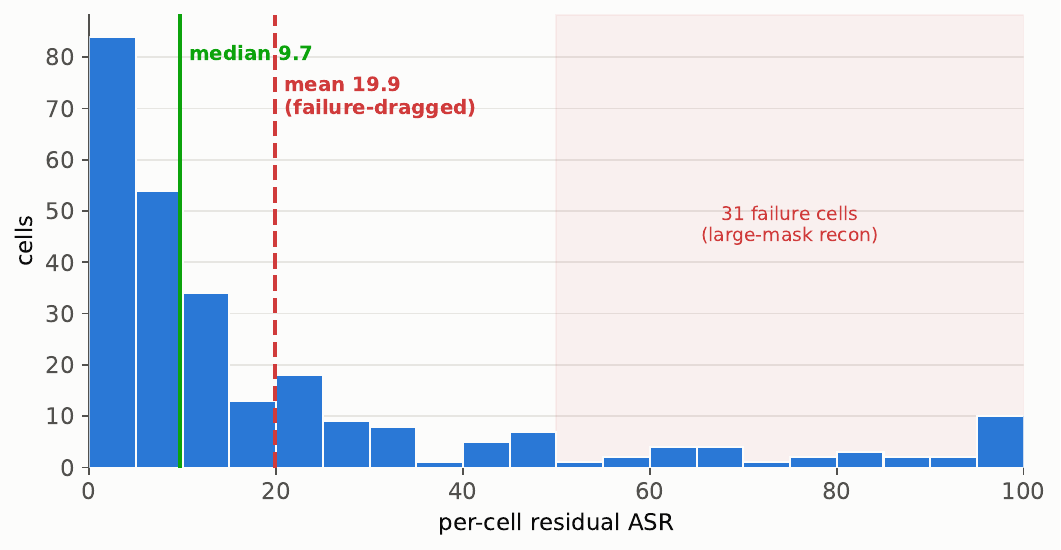}
\caption{Distribution of per-cell residual ASR (bimodal; median $9.7$, small large-mask tail).}
\label{fig:rmhist}
\end{figure}

As we showed in Section~\ref{sec:detection}, the proposed LoRAcle is capable of detecting the class targeted by a backdoor in $86\%+$ of the cases. The most probable classes can then be used to erase the knowledge of the backdoor as explained in Section~\ref{sec:removalMode}.
As can be see in Figure~\ref{fig:rmpareto} and Table~\ref{tab:removal}, our LoRAcle repair holds the best clean accuracy of any method while removing the backdoor. The distillation baselines reach comparable or lower ASR only by collapsing clean accuracy, I-BAU to CDA $54.6$, NAD to $64.5$, while the pruning baselines (ANP, Fine-Pruning) keep accuracy but leave $1.6\times$ more of the backdoor intact ($32$ ASR vs.\ ours $19.9$). No method beats ours on both axes. The removal is reproducible across three seeds; the median seed-to-seed ASR std is $1.0$, and no low-ASR cell blows up on re-run, so the residual is a property of the cell, not run noise. Residual ASR is bimodal (Figure~\ref{fig:rmhist}): most cells sit near zero, median $9.7$, with a small large-mask tail, so LoRAcle drives the ASR close to zero in the large majority of poisoned models while, uniquely, preserving clean accuracy.

\begin{table}[t]\centering\scriptsize
\caption{Backdoor removal results (correctly-detected target).}
\label{tab:removal}
\resizebox{0.9\columnwidth}{!}{%
\begin{tabular}{l c ccc}
\toprule
Method & ASR $\downarrow$ & ASR med.\ $\downarrow$ & CDA $\uparrow$ \\
\midrule
NAD & 12.7 & 9.8 & 64.5 \\
I-BAU & 8.3 & 5.6 & 54.6 \\
Fine-Pruning & 32.6 & 12.8 & 72.6 \\
ANP & 32.2 & 9.3 & 76.9 \\
\textbf{Ours (LoRAcle)} & 19.9 & 9.7 & \textbf{78.2} \\
\bottomrule
\end{tabular}}
\end{table}

The per-dataset, per-trigger breakdown (Table~\ref{tab:removal-granular}) shows the repair reaches low
residual ASR in most cells, with clean accuracy tracking each dataset's clean ceiling. The spread ($\pm$)
localizes the difficulty to a few hard cells (MNIST-blend, CIFAR100-badnets) rather than a broad failure.

\begin{figure}[t]\centering
\begin{subfigure}{\linewidth}\centering
  \includegraphics[width=\linewidth]{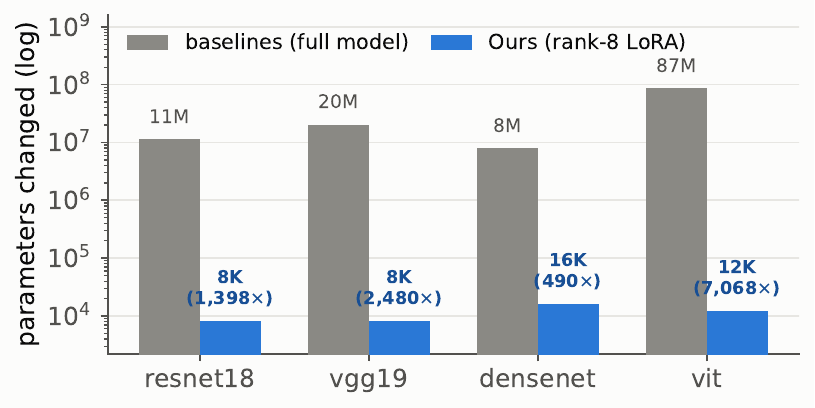}
  \caption{Parameters modified (log scale).}\label{fig:eff-params}\end{subfigure}

\vspace{5pt}
\begin{subfigure}{\linewidth}\centering
  \includegraphics[width=\linewidth]{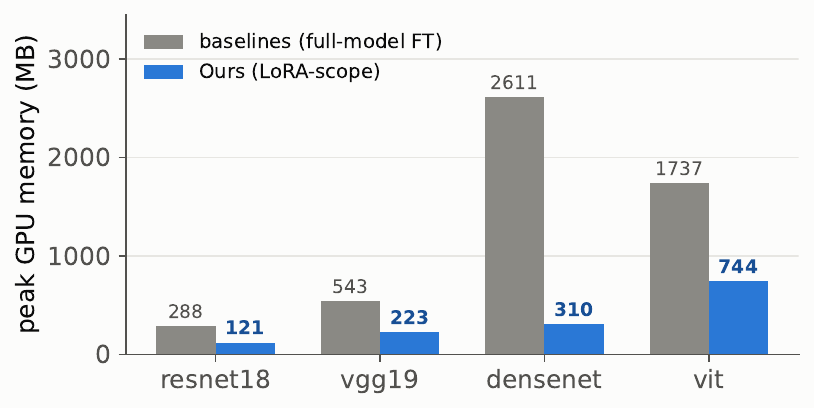}
  \caption{Peak GPU memory (MB).}\label{fig:eff-mem}\end{subfigure}
\caption{Removal efficiency: parameters modified (a) and peak GPU memory (b), LoRAcle vs.\ full-model repair.}
\label{fig:effrm}
\end{figure}

\begin{table}[t]\centering\scriptsize
\caption{Granular backdoor removal by dataset and trigger (correctly-detected target).}
\label{tab:removal-granular}
\resizebox{\columnwidth}{!}{%
\begin{tabular}{ll ccc}
\toprule
Dataset & Trigger & base ASR & residual ASR & CDA\\
\midrule
\multirow{3}{*}{MNIST} & badnets & 79 & 18.5$\pm$15.3 & 97.1$\pm$3.6\\
 & blend & 100 & 48.5$\pm$31.6 & 97.2$\pm$3.3\\
 & wanet & 95 & 10.7$\pm$8.0 & 97.2$\pm$3.4\\
\midrule
\multirow{3}{*}{CIFAR10} & badnets & 99 & 26.8$\pm$36.0 & 72.0$\pm$14.1\\
 & blend & 98 & 13.6$\pm$11.6 & 71.9$\pm$13.5\\
 & wanet & 79 & 31.6$\pm$35.5 & 70.8$\pm$14.6\\
\midrule
\multirow{3}{*}{GTSRB} & badnets & 99 & 7.2$\pm$8.2 & 91.4$\pm$6.3\\
 & blend & 99 & 3.2$\pm$3.9 & 90.3$\pm$5.4\\
 & wanet & 86 & 18.4$\pm$23.3 & 93.8$\pm$4.9\\
\midrule
\multirow{3}{*}{CIFAR100} & badnets & 100 & 46.1$\pm$42.8 & 52.9$\pm$14.1\\
 & blend & 87 & 7.3$\pm$8.8 & 54.7$\pm$13.9\\
 & wanet & 88 & 17.7$\pm$28.4 & 54.6$\pm$13.5\\
\midrule
\multicolumn{2}{l}{\textbf{Overall}} & -- & \textbf{19.9}$\pm$29.0 & \textbf{78.2}$\pm$20.2\\
\bottomrule\end{tabular}}
\end{table}
\textbf{Removal efficiency.} Because detection and repair live in a rank-$r$ subspace, they modify $8$-$16$K parameters versus the $8$-$87$M of full-model defenses, a $490$-$7060\times$ reduction, at $2.3$-$8.4\times$ lower peak GPU memory (Figure~\ref{fig:effrm}). Clean data is matched at $5\%$ of the training set size for every method, so the saving is not bought with more data.

\subsection{Confidence-Adaptive Budget}
\begin{figure}[t]\centering
\includegraphics[width=0.8\linewidth]{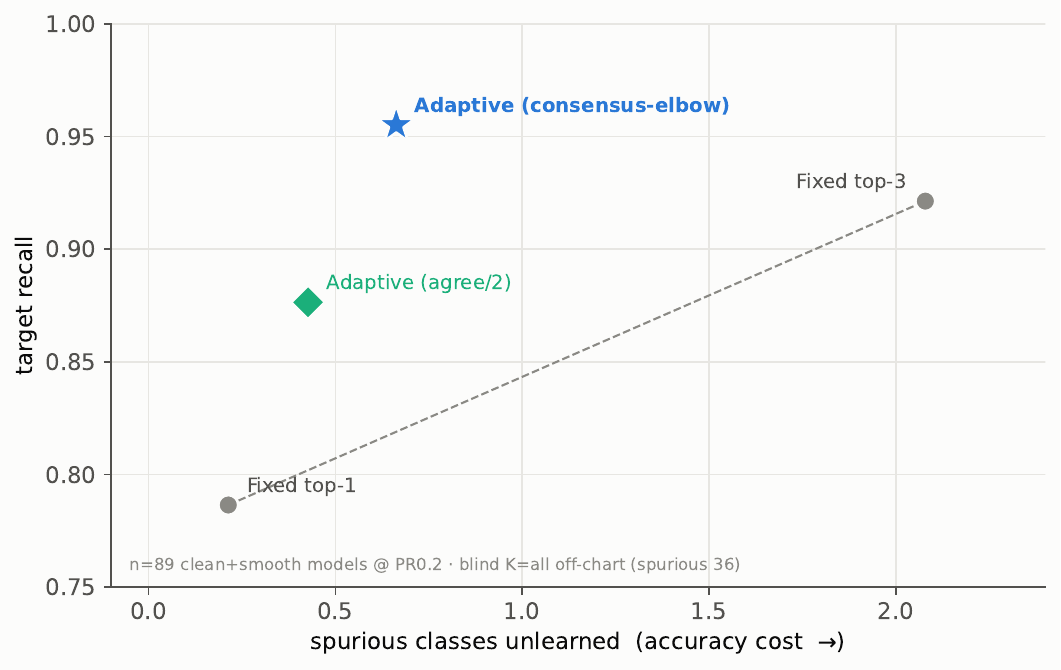}
\caption{Adaptive removal budget: the consensus-elbow policy ($\star$) vs.\ a fixed top-3 budget.}
\label{fig:cluster}
\end{figure}

The consensus of the two probes converts detection confidence into a removal budget
(Fig.~\ref{fig:cluster}). On agreement ($\approx 70\%$ of models) the target is correct $0.98$ of the
time at $K{=}1$; on disagreement the cluster (Eq.~\ref{eq:cluster}) widens. The resulting policy attains
recall $0.955$ while unlearning only $0.66$ spurious classes on average, strictly Pareto-dominating a
fixed top-3 budget ($0.921$ recall at $2.08$ spurious), higher recall at a third of the accuracy cost,
with no tunable threshold.

\subsection{Poisoning Rate Analysis the Budget Regulates Itself}
\label{sec:dose}
Every result so far is at poison rate $0.2$, in this section we test across different poisoning rates $\rho\in\{0.01,0.05,0.1,0.2\}$ over CIFAR-10 victims on every setting and letting the consensus rule set the budget with no per-rate tuning (Table~\ref{tab:dose}).

\begin{table}[t]\centering\scriptsize
\caption{CIFAR-10 under the deployed clean${+}$smooth probe pair, balanced over four architectures $\times$ three triggers ($n{=}24/24/24/120$); the policy is identical at every rate.}
\label{tab:dose}
\begin{tabular}{l cccc}
\toprule
poison rate $\rho$ & $0.01$ & $0.05$ & $0.1$ & $0.2$\\
\midrule
probe agreement & $0.46$ & $0.58$ & $0.54$ & $0.78$\\
accuracy $\mid$ agreement & $0.45$ & $0.79$ & $0.85$ & $\mathbf{0.97}$\\
budget $K$ (classes unlearned) & $3.08$ & $2.29$ & $2.54$ & $\mathbf{1.83}$\\
recall & $0.42$ & $0.79$ & $0.83$ & $\mathbf{0.96}$\\
\bottomrule
\end{tabular}
\end{table}

Two quantities move monotonically with the poisoning level As the backdoor is more powerful, agreement becomes a sharper certificate of correctness ($0.45\!\to\!0.97$) and recall rises with it ($0.42\!\to\!0.96$), while the budget the defender must pay contracts overall ($3.08\!\to\!1.83$ classes). Agreement frequency itself rises from $0.46$ to $0.78$ but not monotonically, it drops at $\rho{=}0.1$, so we report the reliability of agreement, not its rate, as the calibrated signal.
The defender never knows $\rho$, and never has to choose $K$. A fixed budget forces a blind bet, pay for three classes always, and overpay by more than one class exactly where the attack is strongest. The consensus rule instead reads the dose off the model itself and spends accordingly and it widens to $\approx 3$ classes when the shortcut is subtle and contracts to $1.83$ when it is well noticible, recovering $0.96$ of targets at $\rho{=}0.2$ for well under two classes of unlearning. Because each spuriously unlearned class costs roughly two points of clean accuracy (Section~\ref{sec:narcissus}), this contraction is a direct reduction in the accuracy the defender must sacrifice, and it is most precise with a high poisoning rate, and therefore when the attack is most dangerous.
The same table bounds the method honestly. At $\rho{=}0.01$ agreement carries almost no information and end-to-end recall falls to $0.42$, so at one-percent poisoning the shortcut is too faint for either view to localize, and identification is an open regime for us as for the field. What the system does not do is fail confidently; it widens the cluster rather than committing to a wrong class.

\subsection{Behavior on Benign Models}
\label{sec:benign}
Every result so far assumes the audited model is backdoored. In our scenario, an auditor does not know if a model has been poisoned, so the option of a clean model should be considered as a valid outcome.
To test this, we trained a clean model for each of the combinations of the considered datasets and architectures, and ran the full detect-and-repair pipeline on each. 
Looking at Table~\ref{tab:benign}, three findings, together show that the gate is conservative, abstaining on three-quarters of clean models, and that the false detection cause a limited cost in CDA.
The residual false alarms are not spread evenly as they are an intersection between two independent clusters of candidates returned by the consensus rule, so their rate falls as the label space grows, $41.7\%$ on the ten-class datasets but only $8.3\%$ on GTSRB and CIFAR-100. This is the opposite of a liability, the regime where the gate is most trustworthy on benign models (many classes) is exactly the one where a human cannot simply visually check every class.

\begin{table}[t]
\centering
\scriptsize
\caption{The consensus gate on 48 benign models (three seeds $\times$ four architectures per dataset).}
\label{tab:benign}
\begin{tabular}{lccc}
\toprule
Label space &

\shortstack{Gate\\fires} &
\shortstack{Correctly\\abstains} \\
\midrule
10-class (MNIST, CIFAR-10) &  41.7\% & 58.3\% \\
Many-class (GTSRB, CIFAR-100) &  \textbf{8.3\%} & \textbf{91.7\%} \\
\midrule
Overall &  25.0\% & 75.0\% \\
\bottomrule
\end{tabular}
\end{table}

\begin{figure}[t]\centering
\includegraphics[width=0.7\linewidth]{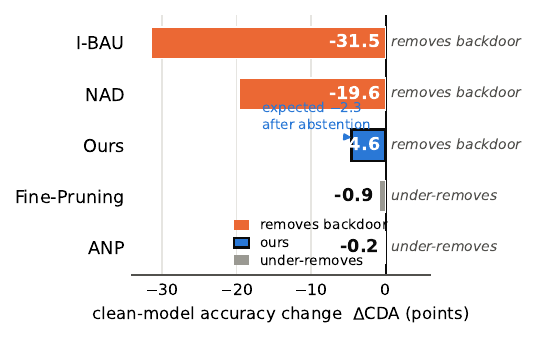}
\caption{Collateral damage on benign CIFAR-10 models.}
\label{fig:collateral}
\end{figure}

When the gate does fire on a clean model, the removal pipeline starts to reconstruct a trigger for the named class and runs a single-class unlearn. Because the repair is guarded by the same clean-accuracy floor used everywhere (keep no checkpoint below base${-}5$), the damage is bounded. Across all $12$ false alarms, the measured clean-accuracy cost is a mean of $2.7$ points and never worse than $4.7$.
Considering a scenario in which we skip the detection phase, and we hand a clean model to repairs unconditionally and pay its full cost every time. Figure~\ref{fig:collateral} measures that cost. The two baselines that erase the backdoor as thoroughly as we do (Section~\ref{sec:results}, Table~\ref{tab:removal}) are precisely the two that destroy a benign model, NAD and I-BAU shed $20$-$31$ accuracy points on average, with one clean model collapsing below $10\%$. The two baselines that leave a benign model intact are the two that under-remove the real backdoor. Ours method removes the backdoor and leaves a healthy model. 

\subsection{Clean-Label Backdoors: Narcissus}
\label{sec:narcissus}
\begin{figure}[t]\centering
\includegraphics[width=\linewidth]{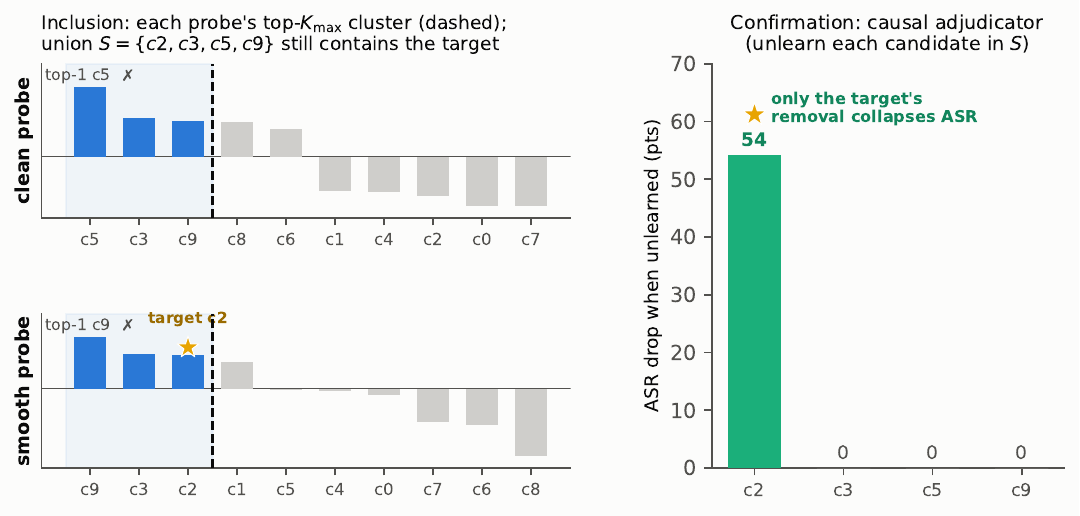}
\caption{Confidence-adaptive recovery on the ResNet-18 Narcissus cell. \emph{Left:} the two probes' scores $P_c$ and the union cluster $S{=}\{c2,c3,c5,c9\}$ (true target c2, $\star$). \emph{Right:} the causal adjudicator, only removing the true target collapses ASR ($54$ vs.\ $0$).}
\label{fig:elbow}
\end{figure}

\begin{figure*}[t]\centering
\begin{subfigure}[t]{0.3\textwidth}\centering
  \includegraphics[width=\linewidth]{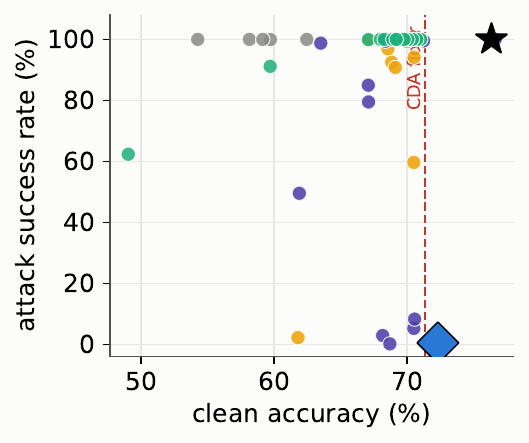}
  \caption{ResNet-18 (base CDA 76.3)}\label{fig:narc-r18}\end{subfigure}
\hfill
\begin{subfigure}[t]{0.3\textwidth}\centering
  \includegraphics[width=\linewidth]{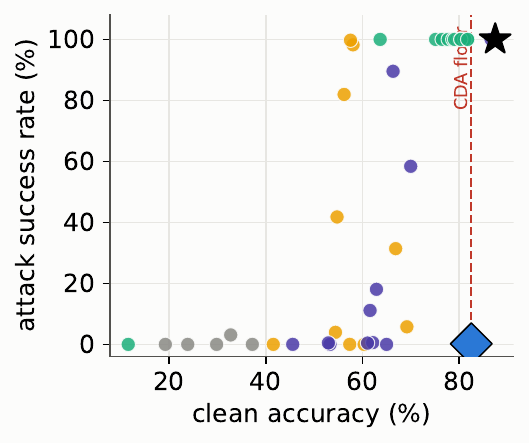}
  \caption{VGG-19 (base CDA 87.4)}\label{fig:narc-vgg}\end{subfigure}
\hfill
\begin{subfigure}[t]{0.3\textwidth}\centering
  \includegraphics[width=\linewidth]{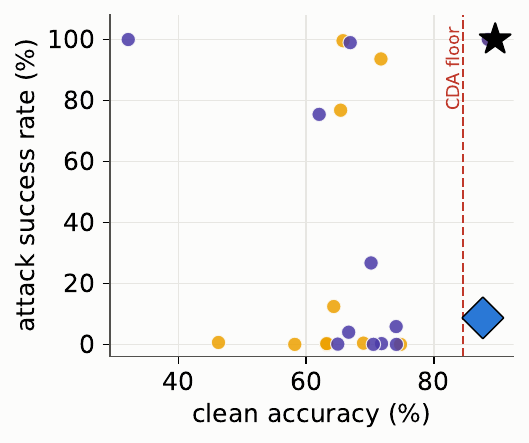}
  \caption{DenseNet (base CDA 89.6)}\label{fig:narc-dense}\end{subfigure}\hfil
\caption{Narcissus removal: clean-accuracy vs.\ ASR per architecture (\emph{down-and-right} is better). $\star$ = poisoned model; $\blacklozenge$ = our LoRAcle erase; other markers = full-model baselines (Fine-Pruning, ANP, NAD, I-BAU); dashed = $-5$ CDA floor. ViT is an adapter-scope boundary (Appendix~\ref{app:narcxdata}).}
\label{fig:narctrade}
\end{figure*}

The attacks considered in the previous experiments are all dirty label attacks; we add a clean-label attack, Narcissus~\cite{zeng2023narcissus}, as the adversary that most stresses a weight-space reader since the trigger is a full-image, low-frequency perturbation optimized so that poisoned target-class images keep their correct label, leaving no dirty-label shortcut to another class to read. Beyond the official CIFAR-10 reference trigger, we generate working clean-label triggers for CIFAR-100 and GTSRB with a cross-dataset out-of-distribution surrogate (Appendix~\ref{app:narcgen}).
On detection, the clean-label attack is the hardest case that defeats every target-identification method (Appendix~\ref{app:narcdetect}). With no dirty-label shortcut, the target class carries no distinctive weight-space signature; it is statistically concealed in the benign space (Appendix~\ref{app:narcdetect}). This limitation is shared across all detectors, and it is precisely why our pipeline is built to degrade gracefully rather than to detect confidently, as the rest of this section shows.
Once the target class is known, our LoRAcle is capable of removing the clean-label backdoor. On CIFAR-10 it drives residual ASR to $0.0$-$8.7$ at a clean-accuracy cost of only $0.1$-$6.8$ points on the three convolutional victims, and to ASR $0$ on ViT once the adapter is extended onto the transformer-MLP subspace the trigger occupies, at a larger $14$-point cost (Table~\ref{tab:narc-removal}, Figure~\ref{fig:narctrade}, Appendix~\ref{app:narcxdata}). The generated CIFAR-100 and GTSRB victims behave the same way (Table~\ref{tab:narc-xdata}). The baselines can also push ASR toward zero, but only by sacrificing $18$-$76$ accuracy points (Figure~\ref{fig:narctrade}): the trigger is a diffuse, full-image pattern, so a full-model prune or fine-tune must overwrite features shared with clean inputs, whereas our rank-8 adapter edits only the target class's shortcut. 
Because this advantage depends on editing the correct shortcut, the right target class, what the pipeline does when point identification fails is decisive, and clean-label is where it fails: on $3$ of the $4$ CIFAR-10 architectures both probes miss the target. The naive response, blindly unlearning all classes, is catastrophic here (Table~\ref{tab:narc-allclass}), every Narcissus class reconstructs a large, diffuse trigger, so forcing one rank-8 adapter to be invariant to all of them either craters clean accuracy ($-22$ CDA on ResNet-18) or exhausts its capacity and under-removes, $-13$ CDA with ASR still $100$ on VGG-19, and under the MLP-extended adapter, ASR $26.6$ on ViT (Table~\ref{tab:narc-allclass}).
This is exactly what the confidence-adaptive budget is built for, and clean-label is its sharpest test. Because the probes disagree, the consensus reads low confidence and widens to the elbow cluster of Eq.~\ref{eq:cluster} (mean size $3$, not all $N$ classes; Figure~\ref{fig:cluster}), which still contains the true target on $3/4$ architectures (Table~\ref{tab:narc-recover}); the causal adjudicator (Section~\ref{sec:removalMode}) then confirms it by removal, on ResNet-18, unlearning the true target drops ASR by $54$ points while all three decoys drop $0$ (Figure~\ref{fig:elbow}). The recovered target is then erased residual ASR $1.55$ (ResNet-18) and $0.09$ (DenseNet) at $72$/$87$ CDA, single-digit, on par with the oracle-target removal of Table~\ref{tab:narc-removal}, yet reached without any point detector, ours included, being able to name the target. The residual case is a honest boundary, on VGG-19 both probes are confidently wrong and the elbow cluster stays too narrow to include the target. The system never emits a confident wrong class; it widens, confirms by causal effect, or abstains.

\begin{table}[t]\centering\scriptsize
\caption{Detection-free recovery on CIFAR-10 Narcissus (true target $=$ class 2), aggregated over all four architectures.}
\label{tab:narc-recover}
\resizebox{\columnwidth}{!}{%
\begin{tabular}{l cc c c cc}
\toprule
Arch & clean & smooth & cluster $S$ & target found & ASR (base$\to$final) & CDA (base$\to$final)\\
\midrule
DenseNet & c0 & c5 & $\{0,1,2,5\}$ & \checkmark & $100\to\mathbf{0.1}$ & $89.6\to87.1$\\
ResNet-18 & c5 & c9 & $\{2,3,5,9\}$ & \checkmark & $100\to\mathbf{1.6}$ & $76.3\to71.9$\\
VGG-19 & c5 & c6 & $\{5,6\}$ & $\times$\ (target $\notin S$) & $100\to76.2$ & $87.4\to84.3$\\
ViT$^\dagger$ & c6 & c2 & $\{2,6\}$ & \checkmark & $100\to\mathbf{0.0}$ & $52.0\to38.0$\\
\bottomrule
\end{tabular}}
\end{table}

\begin{table}[t]\centering\scriptsize
\caption{Narcissus (clean-label) removal with our LoRAcle repair, CIFAR-10.}
\label{tab:narc-removal}
\begin{tabular}{l cc cc}
\toprule
\multirow{2}{*}{Architecture} & \multicolumn{2}{c}{before} & \multicolumn{2}{c}{after repair}\\
 & CDA & ASR & ASR $\downarrow$ & CDA ($\Delta$)\\
\midrule
ResNet-18 & 76.3 & $\approx$100 & \textbf{0.5} & 72.3 ($-$4.0)\\
VGG-19 & 87.4 & $\approx$100 & \textbf{0.2} & 82.5 ($-$4.9)\\
DenseNet & 89.6 & $\approx$100 & \textbf{8.7} & 87.7 ($-$1.9)\\
ViT & 52.0 & $\approx$100 & \textbf{0.0} & 38.0 ($-$14.0)\\
\bottomrule
\end{tabular}
\end{table}

\begin{table}[t]\centering\scriptsize
\caption{Detection-free removal for clean-label. $^\ddagger$ViT uses the MLP-extended adapter (Appendix~\ref{app:narcxdata}).}
\label{tab:narc-allclass}
\begin{tabular}{ll cc}
\toprule
Attack / architecture & budget & $\Delta$CDA & final ASR\\
\midrule
\multirow{2}{*}{badnets (dirty, reference)} & oracle ($K{=}1$) & $-0.1$ & $3$\\
 & all ($K{=}10$) & $-17.4$ & $5$\\
\midrule
\multirow{2}{*}{narcissus / ResNet-18} & oracle & $-2.7$ & $1.1$\\
 & all & $-22.1$ & $0.0$\\
\midrule
\multirow{2}{*}{narcissus / VGG-19} & oracle & $-3.4$ & $15.3$\\
 & all & $-13.4$ & $\mathbf{100.0}$\\
\midrule
\multirow{2}{*}{narcissus / ViT$^\ddagger$} & oracle & $-9.8$ & $0.4$\\
 & all & $-7.5$ & $\mathbf{26.6}$\\
\bottomrule
\end{tabular}
\end{table}

\subsection{Scaling to Larger Label Spaces}

The lens also survives an order-of-magnitude larger label space and a $4\times$ larger input: on Tiny-ImageNet ($200$ classes, $64\times64$) BadNets victims (ResNet-18 and DenseNet, ${\ge}99.8\%$ ASR at $0.1$ poison), the backdoor target is recovered as the most anomalous class of $200$, Top-1 on ResNet-18 and Top-2 on DenseNet, up from rank $180$ and $132$ under the fixed small-scale rule, once the reader is calibrated with the two knobs the lens already exposes: a larger probe rank, and reading energy as a magnitude outlier ($z_C+|z_E|$) rather than the signed $z_C-z_E$, because the sign of the target's energy anomaly inverts with the poisoning proportion. The full analysis, including the member/non-member physics behind the sign inversion, is in Appendix~\ref{app:scale}.

\providecommand{\cmark}{\ensuremath{\checkmark}}
\providecommand{\xmark}{\ensuremath{\times}}

\section{Extended Analysis: Adaptive Attack}
\label{sec:extended}
\begin{figure}[t]\centering
\includegraphics[width=\linewidth]{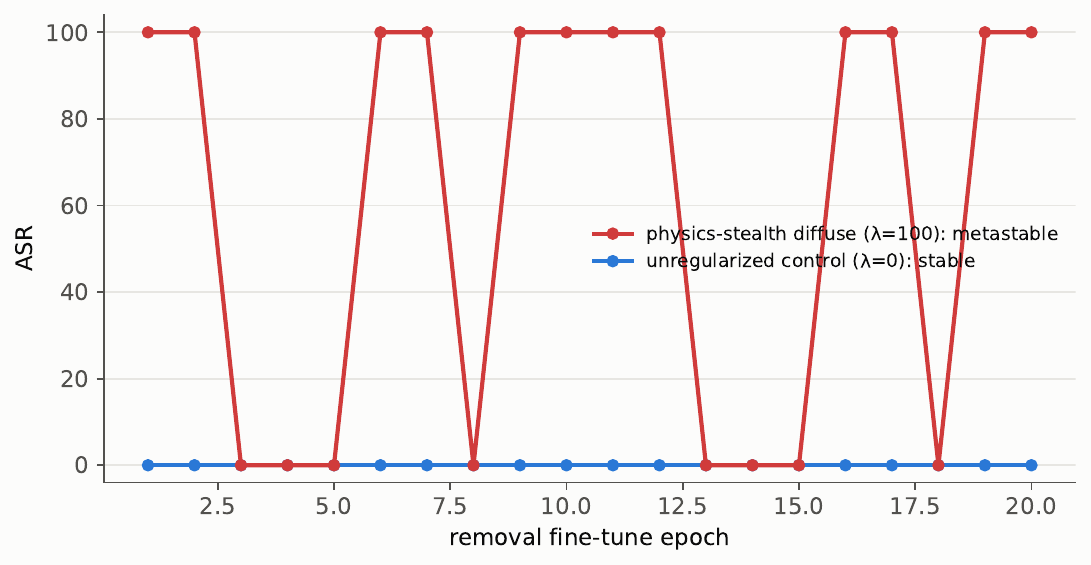}
\caption{Red-team boundary: a physics-stealth full-image trigger's ASR oscillates $0\!\leftrightarrow\!100$ across removal epochs (red) vs.\ a stably-descending control (blue).}
\label{fig:meta}
\end{figure}

As we showed in Section~\ref{sec:dose}, identification of the target class improves monotonically with poison rate, stronger backdoors embed a more legible shortcut, and the adaptive budget contracts as they do. In this direction, we stress-test LoRAcle with an adaptive attacker defined relative to the auditor's threat model of Section~\ref{sec:threatModel}, it knows the full detection mechanism, the two-probe energy/alignment statistic and the rank-$r$ readout, and the target architecture; it knows that the auditor holds some clean reference set but not $D_{\mathrm{clean}}$'s contents; it controls the entire training process, including a stealth regularizer layered onto the poisoning objective; and its only constraint is preserving attack success. It is not told the causal adjudicator or removal procedure, since evasion is scored purely on whether the resulting geometry escapes detection. We trained backdoors with a stealth regularizer that pushes the target's penultimate geometry back toward clean, attacking the exact signal we read, across a range of trigger diffuseness. As a result, we could not produce a backdoor that was simultaneously invisible and unremovable, so to evade reconstruction, the trigger must spread out, and a diffuse trigger is shallow enough for a held-out clean fine-tune to erase.
The one genuinely hard case is a full-image stealth trigger, which becomes metastable under removal; its ASR oscillates between $0$ and $100$ across epochs rather than converging (Figure~\ref{fig:meta}); a removed checkpoint exists and is reliably found, but we report this as ``a good checkpoint is found,'' not as convergence.

The same rank-8 fit already can dump the physics at every layer of the model, not just the penultimate one we use by default. Fusing the penultimate and classifier-head readouts improves target identification on every dataset with no regression, lifting aggregate Top-1 from $86.7$ to $90.9$ (CIFAR-100 reaching $100$; Table~\ref{tab:readout}). Both layers come from the one adapter fit, so parameters, memory, and probe count are unchanged. The one trade-off is on the confidence gate: on the clean models we hold out, the head readout raises spurious two-probe agreement, leading to a false ``high-confidence'' flag from $0$ to ${\sim}2/3$, because the classifier head carries a dominant-class bias that both probes rely on. We therefore keep the penultimate readout as the default and present the head-inclusive fusion as an opt-in accuracy lever; a confidence-preserving multi-layer gate is a concrete, zero-overhead extension.

\section{Limitations and Future Work}
\label{sec:limitations}
The mechanism is strongest on convolutional backbones; on ViT the signature is weaker, and the default rank-8 probe is not uniformly architecture-independent so that the transformer-MLP subspace can require an extended adapter to read and to repair (Appendix~\ref{app:narcxdata}). Detection degrades on from-scratch stealth-trained models, so coverage there rests on removal rather than identification. Removal on diffuse stealth triggers is best-effort checkpoint selection, not convergence. Finally, the granting of a small clean set is a genuine assumption that separates our setting from fully
data-free detectors such as MM-BD.

On target identification, it fits one diagnostic adapter per candidate class, an $O(|\mathcal{Y}|)$ cost; a naive scan of a $1000$-class label space is $1000$ cheap (five-epoch, rank-$8$) fits. This is a practical, not a fundamental, limit: the probe is trivially parallel, and a coarse-to-fine schedule removes the linear factor by ranking classes with one forward pass over the clean set, then running the adapter probe only on the top-$k$ shortlist, using the same consensus/cluster machinery that already sizes the removal budget (Section~\ref{sec:dose}). We evaluate the full scan for faithfulness; the shortlisted variant is the deployable form at scale.

\section{Related Work}
\label{sec:related}

Low-Rank Adapters \cite{hu2022lora} were introduced to enable efficient training of large models by updating only a small subset of parameters, making fine-tuning feasible even on less capable GPUs. LoFT~\cite{fu2024loft} explores the integration of Low-Rank Adaptation into adversarial training, restricting updates to a low-rank parameter space to enable efficient and memory-friendly fine-tuning while maintaining robustness against adversarial perturbations. LoRA can also be efficiently deployed in distributed scenarios, exposing it to different vulnerabilities: FedShield~\cite{mia2025fedshield} combines fully homomorphic encryption with pruning for secure parameter-efficient fine-tuning, restricting updates to LoRA adapter layers.
More in line with this paper, LoRA-Leak~\cite{ran2025lora} assesses LoRA models' vulnerability to MIAs, showing pre-trained models raise privacy risk, while MP-LoRA~\cite{luo2025privacy} trades a proxy attacker's inference gain against the adaptation loss via a min-max adaptation. Although LoRA's MIA security has been studied, to the best of our knowledge, no work has investigated LoRA adapters for explainability, to detect data presence during pre-training.
On the backdoor front, several studies have investigated both attacks on LoRA adapters and defenses. PADBench~\cite{sun2025peftguard} introduced a 13,000-adapter benchmark on which PEFTGuard was proposed as the first PEFT-specific backdoor detector. PEFTGuard targets a different threat model from ours: the adapter itself is the artifact under audit, e.g.\ a LoRA checkpoint shared on a hub, supervised over thousands of labeled benign/backdoored adapters, whereas we audit a fully-trained victim model whose backdoor need not have been introduced through PEFT at all, fitting the adapter ourselves as a diagnostic rather than receiving one to vet, with no labeled adapter corpus to learn from. LoBAM~\cite{yin2024lobam} amplifies malicious weights to reach a high attack success rate at minimal computational cost, showing LoRA-based models remain vulnerable under realistic constraints. To the best of our knowledge, no prior work has fully leveraged the explainability potential of LoRA adapters to extract detailed information about the training data or model behavior.

Post-training backdoor scanning inverts a trigger per class (Neural Cleanse~\cite{wang2019neural}) or
scans classes efficiently (K-Arm~\cite{shen2021backdoor}); feature-space and unified inversions~\cite{WangMZM23} target
input-stealthy attacks. Data-free detectors read the logit landscape (MM-BD~\cite{wang2024mm}) or perturb
neurons (BAN~\cite{xu2024ban}
, a model-level verdict). Sample-level filters (Spectral Signatures, Activation Clustering, ScaleUp)
require the poisoned data. Removal spans distillation (NAD~\cite{lineural}), minimax unlearning
(I-BAU~\cite{zengadversarial}) and pruning (Fine-Pruning~\cite{liu2018fine}, ANP~\cite{wu2021adversarial}). Unlike these, we do not
add a point method: we show a single low-rank lens spans identification, membership and repair, and couple
detection confidence to repair cost.
\section{Conclusion}
\label{sec:conclusion}
We presented \textsc{LoRAcle}, a low-rank lens on what a pretrained model has \emph{internalized}. We
proved the lens on membership, where ground truth is available; recognized a backdoor as malicious
internalization the same lens detects, naming its target class label-free; and erased it in the same
low-rank subspace, with detection confidence sizing the repair. The resulting detect-and-erase pipeline is
competitive with recent detectors, and the most consistent across datasets, at orders-of-magnitude
lower parameter and memory cost, with removal that preserves clean accuracy better than baselines. The
takeaway is not a new defense but a principle: a single, membership-validated low-rank lens can find and
remove latent vulnerabilities in models we did not train.

\bibliographystyle{plain}

\appendix

\section{Implementation details}
\label{app:impl}
\paragraph{Grid} We evaluate four datasets (MNIST, CIFAR-10, GTSRB, CIFAR-100) at $32\times32$ input, four
backbones (ResNet-18, VGG-19, DenseNet, ViT), the dirty-label triggers BadNets, Blend and WaNet, and the
clean-label Narcissus, at poison rates $\{0.01,0.05,0.1,0.2\}$. Detection and removal numbers are reported
at poison rate $0.2$ unless stated; the dose-response (\S\ref{sec:extended}) sweeps all four rates.

\paragraph{\textsc{LoRAcle} probe} For each candidate class we synthesize class proxies by model inversion
($1000$ steps) and fit a rank-$8$ LoRA adapter for at most five epochs (lr $10^{-2}$; $3\times10^{-4}$ for
ViT), then read the adapter ``physics'' (energy, cosine, spectral rank) at the penultimate (\texttt{fc1})
readout. The presence/identification gate fuses two independent probes (a clean and a smooth proxy
generator); agreement yields a confident $K{=}1$ target, disagreement widens to a self-sized cluster.

\paragraph{Removal} Given the target we reconstruct a Neural-Cleanse-style trigger ($1000$ steps, three
restarts, keep the smallest mask), stamp it on $n_{\mathrm{clean}}{=}2000$ held-out clean images with their
\emph{true} labels, and fine-tune \emph{only} the rank-$8$ adapter (unlearn epochs $10$--$25$, lr
$10^{-3}$--$3\times10^{-3}$). All repair methods, ours, Fine-Pruning, ANP, use the identical guarded
best-epoch rule: keep the lowest-ASR checkpoint whose clean accuracy stays within $5$ points of the base.

\paragraph{Baselines} Detection: Neural Cleanse, MM-BD (data-free), K-Arm (clean data $+$ weights).
Removal: Fine-Pruning, ANP, NAD, I-BAU. NAD/I-BAU's fixed forward structure loads only the
ResNet-18/VGG-19 poisoned models, hence their restricted coverage. All experiments run on two GPUs
($\sim\!140$\,GB each) in PyTorch.

\section{Cross-dataset-OOD Narcissus generation}
\label{app:narcgen}
The official Narcissus trigger is a CIFAR-10, target-class-2 universal perturbation and does not transfer to
other datasets (wrong channels/semantics). To obtain a multi-dataset clean-label testbed we \emph{generate}
triggers with a cross-dataset out-of-distribution (OOD) surrogate, using no in-distribution poisoned data
beyond the target class itself. The recipe: (1)~train a surrogate ResNet-18 on a concocted set $=$ an OOD
dataset (all its classes) $+$ the target class as one extra label; (2)~warm up the surrogate on the target
class only; (3)~optimize a trigger $\delta$ \emph{on target-class images} to \emph{maximize the target
logit} (not cross-entropy), constrained to $\lVert\delta\rVert_\infty \le \epsilon = 0.125$ ($32/255$);
(4)~evaluate at $3\times$ test-time magnification. Three design choices are load-bearing, each an ablation:
a cross-entropy objective \emph{saturates} once the target images are already classified and yields a weak
trigger ($\sim\!0.1\%$ ASR); optimizing $\delta$ on \emph{OOD} images gives a ``flip-OOD'' feature that does
not transfer to the victim ($\sim\!2\%$ ASR); optimizing on \emph{target} images with a logit-max objective
fills the $\epsilon$ budget and transfers. Magnitude also matters: $\epsilon=0.063$ gives only $23\%$ ASR,
while $\epsilon=0.125$ (twice the official value) yields a strong backdoor. We use CIFAR-10 as the OOD pool
for both CIFAR-100 and GTSRB. Resulting victim strength is in Table~\ref{tab:narc-gen} (ASR at $3\times$ /
clean accuracy):

\begin{table}[h]\centering\scriptsize
\caption{Generated-victim strength (ASR@$3\times$ / CDA, \%). ViT is arch-fragile: it takes on GTSRB but
fails to internalize the clean-label trigger on CIFAR-100.}
\label{tab:narc-gen}
\begin{tabular}{l cccc}
\toprule
Dataset (OOD pool) & R-18 & VGG-19 & Dense & ViT\\
\midrule
CIFAR-100 (CIFAR-10) & 94.6/63 & 97.4/61 & 94.4/73 & 0.8/20\\
GTSRB (CIFAR-10) & 55/92 & 38/96 & 81/97 & 94/81\\
CIFAR-10 (official ref.) & $\sim$100 & $\sim$100 & $\sim$100 & $\sim$100\\
\bottomrule
\end{tabular}
\end{table}

\section{Why clean-label target-ID fails: three falsified attack-agnostic detectors}
\label{app:narcdetect}
A defender cannot assume an attack is clean-label, so any detector must be attack-agnostic. No target-ID
method, ours or prior work, reliably recovers the clean-label target (Table~\ref{tab:narc-detect}):
reconstruction and trigger-search never fire, the margin statistic MM-BD catches only one architecture, and
our own reader is inconsistent. Below we build three attack-agnostic detectors that each try to exploit
whatever residue clean-label leaves, and report why each fails, the joint evidence that the clean-label
signature is sub-noise.

\begin{table}[h]\centering\scriptsize
\caption{Target identification on Narcissus (CIFAR-10): \emph{no} detector is reliable.
\cmark~=~true-target Top-1, $\circ$~=~Top-3 only, \xmark~=~miss, ---~not evaluated. Reconstruction (Neural
Cleanse) and trigger-search (K-Arm) never fire; MM-BD catches only ResNet-18; our reader reaches Top-3 on
three backbones but on a single fixed probe that is probe-unstable (finding~C).}
\label{tab:narc-detect}
\begin{tabular}{l cccc c}
\toprule
Detector & R-18 & VGG-19 & Dense & ViT & Top-1\\
\midrule
Neural Cleanse & \xmark & \xmark & --- & --- & $0/2$\\
K-Arm & \xmark & \xmark & \xmark & \xmark & $0/4$\\
MM-BD & \cmark & \xmark & \xmark & $\circ$ & $1/4$\\
\textsc{LoRAcle} (ours) & $\circ$ & \xmark & $\circ$ & \cmark & $1/4$\\
\bottomrule
\end{tabular}
\end{table}

\paragraph{(A) Reconstruction cost} Since removal reconstructs the target's trigger, we ran Neural-Cleanse
reconstruction for \emph{every} class and ranked by mask size. On narcissus victims every class reconstructs
a compact trigger ($\sim\!30$--$50$ px of $1024$) at $\sim\!96\%$ ASR, and the true target is \emph{not}
anomalously small: rank $6/10$ on ResNet-18, $9/10$ on VGG-19, with a MAD outlier score $\le 0$ (Neural
Cleanse flags outliers above $+2$). Reconstruction is a good remover once the class is known, but
carries no identification signal, every class has a cheap trigger.

\paragraph{(B) Two-sided cosine anomaly} A dirty-label target carves a \emph{high, positive} adapter-cosine
anomaly; the clean-label target sits on the \emph{opposite} (negative) tail, so the signed detector score
ranks it last while its magnitude $|z|$ is $2$nd--$3$rd largest on a single probe. But converting the ranker
to $|z|$ recovers narcissus only by destroying everything else (Table~\ref{tab:narc-twosided}): dirty-label
Top-1 collapses and the clean-model presence gate \emph{inverts} (clean models outscore backdoored ones).
The target is not the clearest outlier even within its own model, and a clean model produces a larger
negative-tail chance outlier ($|z|\,2.08$) than the real narcissus signal ($|z|\,0.8$--$1.4$): signal-to-noise
$<1$.

\begin{table}[h]\centering\scriptsize
\caption{Two-sided cosine cannot replace the signed rule. Top-1/Top-3 target-ID per attack, and the presence
margin (backdoored minus clean top-score; negative $=$ clean models look more suspicious). Recovering
narcissus with $|z|$ craters dirty-label and inverts the gate.}
\label{tab:narc-twosided}
\resizebox{\columnwidth}{!}{%
\begin{tabular}{l cccc c}
\toprule
Ranker & badnets & blend & wanet & narcissus & presence\\
\midrule
signed ($z_{\cos}\!-\!z_{\mathrm{en}}$) & 86/100 & 71/71 & 57/71 & 0/0 & $+0.19$\\
$|z_{\cos}|$ (two-sided) & 29/57 & 14/71 & 71/100 & 0/100 & $-0.15$\\
$|z_{\cos}|\!-\!z_{\mathrm{en}}$ & 43/86 & 57/71 & 43/71 & 50/50 & $+0.50$\\
\bottomrule
\end{tabular}}
\end{table}

\paragraph{(C) Cross-strategy consistency} A genuine signature should be stable across independent proxy
generators; a chance outlier should scatter. The narcissus target's cosine $z$ \emph{flips sign} across the
sparse/smooth/hybrid generators (e.g.\ VGG-19: $-1.02,+0.74,-0.57$), so averaging, the standard way to lift
a signal above noise, collapses it to $\sim\!-0.28$ ($|z|$-rank $8/10$), while a clean model retains
a mean $|z|$ of $\sim\!1.5$. The apparent anomaly is probe noise, not signature. Together, (A)--(C) show the
clean-label footprint is statistically inside the benign manifold on every axis we can read without an
attack-type oracle.

\section{Cross-dataset narcissus removal, per cell}
\label{app:narcxdata}
Table~\ref{tab:narc-xdata} gives the per-cell LoRAcle removal on the generated CIFAR-100/GTSRB victims
(three seeds each, oracle target, guarded best-epoch rule). Removal drives ASR to $\le 3.7\%$ everywhere,
at $-6.8$ to $+0.2$ CDA; GTSRB is essentially free, CIFAR-100 (a low clean-accuracy victim) costs $\sim\!6$
points. ViT is excluded: its deployed adapter (attention $Q/V$ plus \texttt{fc1}) does not span the
transformer-MLP subspace the clean-label trigger occupies, so the deployed \texttt{fc1}-only unlearn cannot reach it. \rev{Extending the same rank-8 adapter onto the transformer-MLP projections (\texttt{intermediate.dense} and \texttt{output.dense}) does erase it: ASR $100\to0$ at $52\to38$ CDA and only $+737$K adapter parameters, versus the $86$M a full-model unlearn touches (which reaches the same ASR but collapses CDA to $22$). Extending onto the attention-output projection alone does not suffice (ASR stalls at $66$), so the clean-label trigger lives specifically in the MLP subspace. This confirms a subspace-coverage boundary, not an unremovable backdoor.}

\begin{table}[h]\centering\scriptsize
\caption{Cross-dataset narcissus removal (mean$\pm$std over three seeds), three healthy architectures.
Generated triggers (Appendix~\ref{app:narcgen}); oracle target.}
\label{tab:narc-xdata}
\begin{tabular}{ll cc cc}
\toprule
\multirow{2}{*}{Dataset} & \multirow{2}{*}{Arch} & \multicolumn{2}{c}{before} & \multicolumn{2}{c}{after}\\
 & & CDA & ASR & CDA & ASR\\
\midrule
\multirow{3}{*}{CIFAR-100} & ResNet-18 & 63.1 & 94.6 & 57.0$\pm$0.6 & 0.9$\pm$0.9\\
 & VGG-19 & 60.6 & 97.4 & 53.8$\pm$0.0 & 3.7$\pm$3.8\\
 & DenseNet & 73.4 & 94.4 & 67.8$\pm$0.5 & 2.2$\pm$3.7\\
\midrule
\multirow{3}{*}{GTSRB} & ResNet-18 & 92.2 & 55.3 & 90.7$\pm$0.4 & 0.0$\pm$0.0\\
 & VGG-19 & 96.1 & 38.3 & 94.8$\pm$0.1 & 0.0$\pm$0.0\\
 & DenseNet & 97.0 & 80.6 & 97.2$\pm$0.5 & 0.0$\pm$0.0\\
\bottomrule
\end{tabular}
\end{table}

\section{Per-cell backdoor removal}
\label{app:removal-percell}
\rev{Table~\ref{tab:removal-percell} decomposes the aggregate removal result of Table~\ref{tab:removal} into all
$48$ dirty-label cells (four datasets $\times$ four architectures $\times$ \{BadNets, Blend, WaNet\}, per-cell
mean over replicate seeds and targets). The surgical rank-$8$ adapter drives ASR to single digits in the large
majority of cells at a mean clean-accuracy change of $+0.9$ points, reproducing the Table~\ref{tab:removal}
headline (mean residual ASR $19.6$, mean CDA $79.1$). The high-residual entries are exactly the large-mask
tail of Figure~\ref{fig:rmhist}, concentrated on DenseNet and on the diffuse Blend/WaNet triggers where
reconstruction returns no compact operator; these are the cells the mask-gated router
(Section~\ref{sec:removalMode}) defers to the held-out clean fine-tune rather than a surgical edit.}

\begin{table}[h]\centering\scriptsize
\caption{Per-cell surgical removal (ASR base$\to$after, \%; $\Delta$CDA is the mean clean-accuracy change over
the three triggers). Bold marks residual ASR $<10$. This is the per-cell decomposition behind
Table~\ref{tab:removal}; the large-residual cells are the large-mask tail (Figure~\ref{fig:rmhist}) that the
mask-gated router (Section~\ref{sec:removalMode}) defers to clean fine-tune.}
\label{tab:removal-percell}
\resizebox{\columnwidth}{!}{%
\begin{tabular}{ll ccc c}
\toprule
Dataset & Arch & BadNets & Blend & WaNet & $\Delta$CDA\\
\midrule
\multirow{4}{*}{MNIST}
 & ResNet-18 & 100$\to$15 & 100$\to$29 & 99$\to$\textbf{6} & $-0.1$\\
 & VGG-19 & 100$\to$39 & 100$\to$43 & 100$\to$\textbf{6} & $-0.6$\\
 & DenseNet & 100$\to$13 & 100$\to$55 & 100$\to$21 & $0.0$\\
 & ViT & 15$\to$\textbf{10} & 100$\to$44 & 79$\to$15 & $+1.3$\\
\midrule
\multirow{4}{*}{CIFAR-10}
 & ResNet-18 & 98$\to$13 & 96$\to$\textbf{7} & 87$\to$\textbf{7} & $+0.7$\\
 & VGG-19 & 99$\to$34 & 99$\to$13 & 95$\to$13 & $-2.6$\\
 & DenseNet & 99$\to$24 & 99$\to$30 & 98$\to$92 & $+0.6$\\
 & ViT & 98$\to$31 & 98$\to$\textbf{7} & 30$\to$13 & $+0.9$\\
\midrule
\multirow{4}{*}{CIFAR-100}
 & ResNet-18 & 100$\to$17 & 80$\to$\textbf{1} & 90$\to$\textbf{1} & $-0.7$\\
 & VGG-19 & 100$\to$38 & 68$\to$\textbf{2} & 95$\to$\textbf{5} & $-1.2$\\
 & DenseNet & 100$\to$74 & 84$\to$21 & 97$\to$57 & $-2.5$\\
 & ViT & 98$\to$14 & 94$\to$\textbf{3} & 60$\to$\textbf{3} & $+4.2$\\
\midrule
\multirow{4}{*}{GTSRB}
 & ResNet-18 & 99$\to$\textbf{3} & 97$\to$\textbf{0} & 72$\to$\textbf{4} & $+2.5$\\
 & VGG-19 & 99$\to$\textbf{8} & 99$\to$14 & 96$\to$\textbf{3} & $+0.9$\\
 & DenseNet & 99$\to$\textbf{4} & 100$\to$13 & 99$\to$49 & $+1.8$\\
 & ViT & 99$\to$16 & 99$\to$\textbf{7} & 37$\to$\textbf{7} & $+9.1$\\
\bottomrule
\end{tabular}}
\end{table}

\section{Untrained-model control: internalization vs.\ optimization difficulty}
\label{app:untrained-control}
\rev{To isolate internalization from the mere difficulty of the fit, we repeat the familiarity experiment
(Section~\ref{sec:continuum}) on two models of the \emph{same} architecture: the trained model, and a
\emph{random-init} model built by the identical constructor with no checkpoint loaded. The optimization
problem is byte-for-byte identical, same architecture, same rank-$8$ probe, same data, same schedule,
only the internalized structure is removed. Table~\ref{tab:untrained} gives the adapter energy
($\lVert\Delta W\rVert/\lVert W\rVert$) to adapt toward a class for four batch types, and the
familiar-versus-unfamiliar separation AUROC. On the trained model the internalization ordering holds:
in-distribution content is cheaper to adapt toward than out-of-distribution content or Gaussian noise
(AUROC $0.83$ on ResNet-18, $0.99$ on VGG-19). On the random-init model the ordering \emph{collapses}, all
four batch types fall within measurement noise (spread $<0.0003$ on ResNet-18), the separation drops to
chance ($0.21$ / $0.25$), and on VGG-19 the cheapest batch is in fact Gaussian noise, the opposite of an
internalization signal. The absolute energy is smaller on the untrained model (its random weights admit
smaller relative updates), but the diagnostic is the scale-free within-model \emph{ordering}, which requires
internalization to appear. This is the sharpest evidence that the lens reads internalized structure rather
than a correlated property of the optimization landscape.}

\begin{table}[h]\centering\scriptsize
\caption{Untrained-model control. Adapter energy $\lVert\Delta W\rVert/\lVert W\rVert$ to adapt toward a
class, by batch type, on a trained vs.\ a random-init model of the same architecture (mean over $10$ batches
of $128$). The familiar$<$unfamiliar ordering that separates content at AUROC $0.83$--$0.99$ on trained
models vanishes on the random-init models, though the optimization problem is identical.}
\label{tab:untrained}
\begin{tabular}{ll cccc c}
\toprule
Arch & Model & train & unseen & OOD & noise & AUROC\\
\midrule
\multirow{2}{*}{ResNet-18} & trained & 0.051 & 0.049 & 0.053 & 0.054 & \textbf{0.83}\\
 & random & 0.0073 & 0.0074 & 0.0073 & 0.0071 & 0.21\\
\midrule
\multirow{2}{*}{VGG-19} & trained & 0.098 & 0.097 & 0.115 & 0.115 & \textbf{0.99}\\
 & random & 0.017 & 0.017 & 0.017 & 0.010 & 0.25\\
\bottomrule
\end{tabular}
\end{table}

\section{Scaling to larger label spaces}
\label{app:scale}
To test whether the lens survives an order-of-magnitude larger label space and a $4\times$ larger input, we train BadNets victims on Tiny-ImageNet ($200$ classes, $64\times64$) on two backbones. The attack scales cleanly, ResNet-18 and DenseNet both reach ${\ge}99.8\%$ ASR at $0.1$ poison, and so does the internalization signature, but reading it at scale exercises exactly the two levers this paper has already isolated: probe rank (Section~\ref{sec:rankabl}) and readout choice (above). Two things change with scale, both predictable. First, the sign of the target's energy anomaly inverts, and the cause is the same internalization physics we validated on membership (Section~\ref{sec:analysis}): a member batch, content the model has absorbed, given a small, well-aligned, low-energy adapter update, whereas a non-member batch forces a large, high-energy one. The backdoor target is always an energy outlier; which side it lands on is set by how heavily it is loaded, i.e.\ by the poisoning proportion relative to a uniform class share $1/K$. At $32\times32$ ($K{=}10$, poison $0.2$) the target carries only a modest excess, $\rho+(1{-}\rho)/K\approx 0.28$, about $3\times$ a uniform share, of a single, clean, over-learned shortcut; the model internalizes it compactly, so it reads member-like and low-energy, and $z_C-z_E$ correctly flags it. At $200$ classes the same poison, all funnelled into one of $200$ labels, forces the target to absorb ${\sim}21\times$ its share ($0.1{+}0.9/200$ vs.\ $1/200$), a heterogeneous pile the model cannot represent compactly, so adapting toward it now costs a large update and it reads non-member-like and high-energy ($z_E{=}{+}5.3$ ResNet, ${+}2.1$ DenseNet). The anomaly's magnitude is large in both regimes; only its sign tracks the load.
Looking at results in Table~\ref{tab:scale}, the fixed $32\times32$ rule $z_C-z_E$ therefore subtracts the very signal it should add and buries the target (rank $132$-$180$ of $200$); reading energy as a magnitude outlier, $z_C+|z_E|$, tracks the load regime and recovers it. Second, the rank needed to expose the signal grows with the backbone: ResNet-18 identifies the target Top-1 already at rank $8$, whereas DenseNet's feature reuse requires a rank-$64$ probe, the same ``signal strengthens with rank'' behaviour of Section~\ref{sec:rankabl}, now at $200$ classes (DenseNet target rank under $z_C+|z_E|$: $52,14,2,2$ at rank $8,32,64,128$). At a common rank-$64$ probe the scale-adapted readout identifies the target Top-1 (ResNet-18) / Top-2 (DenseNet) of $200$. This is a regime calibration, not a free universal rule: $z_C+|z_E|$ costs ${\sim}8$ Top-1 points back on the $32\times32$ grid ($86.7\!\to\!78.3$), where the target is genuinely low-energy, and the signed rule is correct. The takeaway is that the mechanism transfers, on both architectures, the backdoor target is the single most anomalous class of $200$, once the reader is calibrated to the regime through the same rank and readout knobs the lens already exposes.

\begin{table}[h]\centering\scriptsize
\caption{Tiny-ImageNet ($200$ classes, $64\times64$), BadNets, target class $0$,
$0.1$ poison, rank-$64$ diagnostic probe.}
\label{tab:scale}
\begin{tabular}{l ccc}
\toprule
Backbone & ASR & $z_C-z_E$ (rank) & $z_C+|z_E|$ (rank)\\
\midrule
ResNet-18 & 99.99 & 180 & \textbf{1}\\
DenseNet  & 99.79 & 132 & \textbf{2}\\
\bottomrule
\end{tabular}
\end{table}

\section{Detection efficiency}
\label{app:effdet}
\begin{figure}[h]\centering
\includegraphics[width=0.92\linewidth]{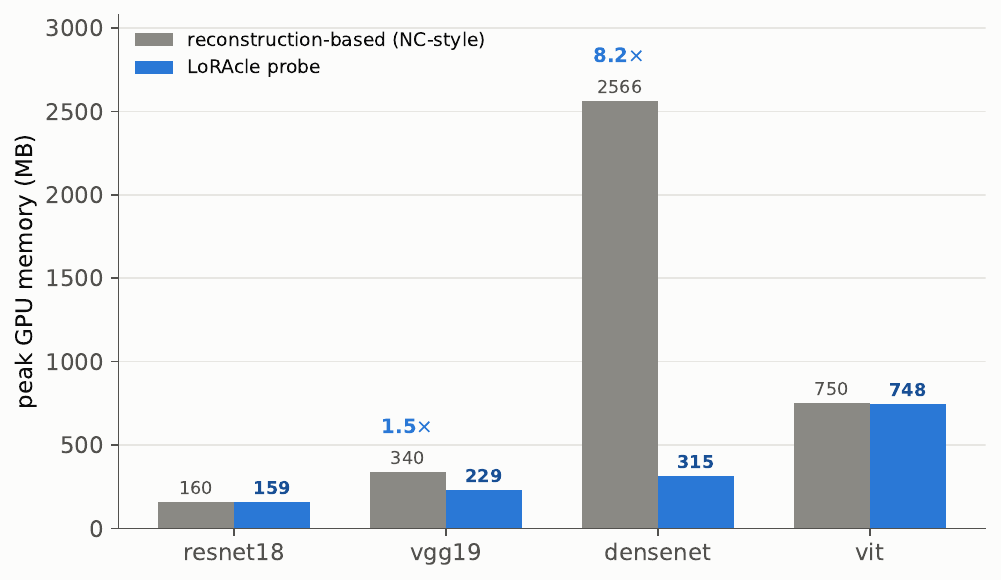}
\caption{Detection memory is architecture-dependent: the probe avoids the input-space backprop that
makes reconstruction costly on DenseNet.}
\label{fig:effdet}
\end{figure}

Figure~\ref{fig:effdet} details the architecture-dependent detection-memory gap discussed in Section~\ref{sec:detection}.

\section{Multi-layer readout fusion}
\label{app:readout}
Table~\ref{tab:readout} reports the classifier-head readout fusion discussed in Section~\ref{sec:extended}: fusing the penultimate and classifier-head readouts from the same adapter fit lifts aggregate Top-1 from $86.7$ to $90.9$ at no extra probe cost.

\begin{table}[h]\centering\scriptsize
\caption{The classifier-head readout improves accuracy. C10=CIFAR10, C100=CIFAR100. Feat.=Penultimate, Feat.$+$Head=Penultimate $+$ classifier.}
\label{tab:readout}
\begin{tabular}{lccccc}
\toprule
Readout & C10 & MNIST & GTSRB & C100 & Avg.\\
\midrule
Feat. & 89.2 & 85.8 & 85 & 86.7 & 86.7\\
Feat.+Head & 90.8 & 89.2 & 89.2 & 100 & \textbf{90.9}\\
\bottomrule
\end{tabular}
\end{table}

\end{document}